\documentstyle[12pt]{article}

\global\arraycolsep=1pt
\begin{document}

\hfill    SISSA/ISAS 73/93/EP

\hfill    June 1993

\begin{center}
\vspace{24pt}
{\large \bf TOPOLOGICAL $\sigma$-MODELS IN FOUR DIMENSIONS\\
AND TRIHOLOMORPHIC MAPS \footnotemark\footnotetext{
Partially supported by EEC, Science Project SC1$^{*}$-CT92-0789}}
\vspace{24pt}

{\sl Damiano Anselmi and Pietro Fr\`e}

\vspace{6pt}

SISSA - International School for Advanced Studies, via Beirut 2-4,
I-34100 Trieste, Italy\\
and I.N.F.N. - Sezione di Trieste, Trieste, Italy\\

\vspace{12pt}

\end{center}

\vspace{24pt}

\begin{center}
{\bf Abstract}
\end{center}

\vspace{12pt}

\noindent
It is well-known that
topological $\sigma$-models in two dimensions
constitute
a path-integral approach to the study of holomorphic maps from a Riemann
surface $\Sigma$ to an almost complex manifold ${\cal K}$,
the most interesting case being that where
${\cal K}$ is a K\"ahler manifold. We show that, in the same way, topological
$\sigma$-models in four dimensions introduce a path integral
approach to the
study of triholomorphic maps
$q:{\cal M} \rightarrow {\cal N} $ between a four dimensional
Riemannian manifold ${\cal M}$
and an almost quaternionic manifold ${\cal N}$.
The most interesting cases are those where ${\cal M},{\cal N}$
are hyperK\"ahler or quaternionic K\"ahler.
BRST-cohomology
translates into intersection theory in the moduli-space of this new
class of instantonic maps, that are named by us hyperinstantons. The definition
of triholomorphicity that we propose is expressed by the equation
$q_*-J_{u}\circ q_* \circ j_u =0$, where $\{j_u,u=1,2,3\}$ is an
almost quaternionic structure on ${\cal M}$ and
$\{J_u,u=1,2,3\}$ is an almost quaternionic structure
on ${\cal N}$. This is a generalization of the Cauchy-Fueter equations.
For ${\cal M},{\cal N}$ hyperK\"ahler,
this generalization naturally arises  by
obtaining the topological  $\sigma$-model as a twisted version of the
N=2 globally supersymmetric $\sigma$-model.
We discuss various examples of hyperinstantons,
in particular on the torus and the K3 surface. We also analyse the coupling
of the topological $\sigma$-model to topological gravity. The classification
of triholomorphic maps and the analysis of their moduli-space is a new
and fully open mathematical problem that we believe deserves the attention
of both  mathematicians and physicists.
\vfill
\eject
\section{Introduction}
\label{intro}

A $\sigma$-model is a theory of maps $\phi:{\cal M}\rightarrow {\cal N}$
from a Riemannian manifold
${\cal M}$ to a Riemannian manifold ${\cal N}$, described by the action
\begin{equation}
{\cal S}={1\over 2}
\int_{\cal M}d^mx\,\sqrt{g(x)}g^{\mu\nu}(x)\partial_\mu\phi^i(x)
\partial_\nu\phi^j(x)h_{ij}(\phi(x)),
\label{saction}
\end{equation}
where $m={\rm dim}\,{\cal M}$, $x$ denote the points of ${\cal M}$, while
$g_{\mu\nu}(x)$ is the metric of ${\cal M}$ and $h_{ij}(\phi)$ is the metric
of ${\cal N}$. In general, ${\cal M}$ is called the world-manifold;
it is called the world-sheet when it is two-dimensional.
${\cal N}$ is the target manifold.

The action ${\cal S}$ is invariant under those infinitesimal deformations
$\phi+\delta\phi$ of the map $\phi$ that are isometries  of ${\cal N}$.
If the ${\cal M}$  metric $g_{\mu\nu}$ is a given background metric,
then we say that gravity is external. One can
also consider the case in which gravity is dynamical. In this case
${\cal M}$ can be arbitrarily chosen only as a topological space:
its metric
$g_{\mu\nu}$, instead, is
to be determined consistently with the map $\phi$,
from the variation of
an action that is the sum of the $\sigma$-model action (\ref{saction})
plus the Einstein-Hilbert action
\begin{equation}
-{1\over 2}\int_{\cal M}d^mx\, \sqrt{g(x)}R(x),
\end{equation}
$R$ being the scalar curvature of $g_{\mu\nu}$.

A topological $\sigma$-model \cite{witten2,besi2,witten3,topsmodel}
is a theory dealing with the
homotopy classes of the maps
$\phi:{\cal M}\rightarrow {\cal N}$. It is described by an action which is
invariant under any continuous deformation
$\phi\rightarrow \phi+\delta\phi$ of the map. This is intrinsically a quantum
field theory, since the classical action is either zero or a topological
invariant, due to the large symmetry that it is required to possess.
Indeed, the functional integral formulation of quantum field theory
provides very powerful methods for the study of
such a theory of maps.
This large symmetry is BRST-quantized
\cite{baulieuYM,topYM} in the usual
ways as any other gauge symmetry
and the gauge is fixed by choosing suitable representatives in the
homotopy classes of the maps $\phi$. These representatives are usually
some kind of instantons, because it is in this case that the topological
theory turns out to be most interesting. The theory is not independent
of the chosen gauge-fixing. Indeed, topological field theories
\cite{witten2,besi2,witten3,topsmodel,baulieuYM,topYM,birmingham,topFT,topG,topG4}
show very clearly that in general two gauge-conditions that are not
continuously
deformable one into the other give rise to inequivalent quantum field theories.
Typically, the gauge-fixing does
not fix the gauge completely, namely there is
a subset of continuous deformations
$\phi\rightarrow \phi+\delta\phi$ that preserve the gauge condition.
The set of maps that satisfy
the instantonic equations is called the moduli-space
of the theory and in general it is a finite dimensional manifold.
The topological
field theory is a cohomological theory in this space.
Indeed, the functional integral is projected onto an integral over the
moduli-space
(that is an ordinary integral) and the physical amplitudes are suitable
topological invariants of this space.

In the case of external gravity there is an alternative definition of
topological field theory, that is a theory characterized by a BRST-exact
energy-momentum tensor $T_{\mu\nu}=(1/\sqrt{g})\delta {\cal S}/\delta
g^{\mu\nu}$.  This is so, because the lagrangian formulation makes explicit
use of the world-metric $g_{\mu\nu}$ although a topological field theory is
expected to describe quantities that are independent of any metric.
This feature is guaranteed by the very BRST-exactness of $T_{\mu\nu}$.
In the case of dynamical gravity, independence
of the metric is a meaningless requirement,
because the metric is a quantum field.
Nevertheless, one can define the concept of topological
gravity by saying that it is a theory that quantizes the most general
continuous deformation of the metric \cite{witten3,topG}.

We see that topological field theories represent a beautiful joint-venture
for physics and mathematics. One conveniently
formulates a mathematical problem
in the language of physics and moreover the object of the study is
of physical interest, since instantons are peculiar solutions to
the field equations, namely solutions
that give a leading contribution to the functional
integral (in the Euclidean region).

Topological $\sigma$-models have been extensively studied in two
dimensions ($m=2$)
\cite{witten2,besi2,witten3,topsmodel,topS}.
The world-sheet $\Sigma$ is a Riemann surface and
the target manifold ${\cal N}$ is almost complex. The instanton
equations are the Cauchy-Riemann equations
\begin{equation}
\partial_\alpha\phi^i-{\varepsilon_\alpha}^\beta {J_j}^i\partial_\beta\phi^j=0,
\label{cauchyriemann}
\end{equation}
where $\alpha=1,2$ labels the world-sheet coordinates,
$\varepsilon$ is the world-sheet complex structure and $J$
is the almost complex structure of ${\cal N}$. Witten showed \cite{witten2}
that a convenient starting point is provided by choosing ${\cal N}$ to be
a K\"ahler
manifold. The theory then describes the holomorphic embeddings
$\phi:\Sigma\rightarrow {\cal N}$ of Riemann surfaces into
K\"ahler manifolds. If $K$ is the K\"ahler form of ${\cal N}$, one
can start from the classical action \cite{besi2}
\begin{equation}
{\cal S}_{class}=\int_\Sigma\phi^*K,
\label{sclass}
\end{equation}
where $\phi^*K$ denotes the pull-back of $K$ onto the world-sheet $\Sigma$.
The action (\ref{sclass}) is clearly
invariant under any continuous deformation of the map
$\phi$.

One reason why a K\"ahlerian ${\cal N}$ is convenient comes once more
from physics. Indeed, ${\cal N}$=K\"ahler manifold
is the condition for a two-dimensional
$\sigma$-model to have an N=2 supersymmetry \cite{N=2sm}. In that case, the
topological $\sigma$-model arises naturally from the N=2 supersymmetric
$\sigma$-model \cite{N=2tw}, by performing a set of formal manipulations
and redefinitions that is called the topological twist \cite{witten,witten2}.
The topological theory that comes from the twist is already gauge-fixed
and the natural gauge-fixing is precisely the instantonic condition.

As a matter of fact, the twist was introduced in four dimensions,
when Witten formulated topological Yang-Mills theory \cite{witten} and
showed that it can be obtained by twisting N=2 super Yang-Mills theory.
In general, although the starting N=2 theory is defined on a flat
world-manifold
${\cal M}$ (when supersymmetry is global),
yet the final theory can be defined on any
${\cal M}$ and it is independent of the choice of
the metric on ${\cal M}$.
Recently \cite{anselmifre,bardonecchia,anselmifre2,erice},
we have shown that any N=2
globally but also locally supersymmetric theory in four dimensions
can be twisted, if the
topological twist is suitably improved. One gets a topological version
of the theories, namely topological gravity, topological Yang-Mills
theory and topological $\sigma$-model, eventually coupled together.
In pure topological gravity \cite{anselmifre,bardonecchia}
or topological gravity coupled to topological
Yang-Mills theory \cite{anselmifre2,erice},
the gravitational instantons \cite{gravinst} are the solutions to the condition
of selfduality for the spin connection $\omega^{ab}$, namely
$\omega^{-ab}=0$. On the other hand,
topological Yang-Mills theory is gauge-fixed by
the equations of the usual
Yang-Mills instantons.
The novelty comes from the twist of the N=2
$\sigma$-model
\cite{anselmifre2,erice}
describing the self-interaction of hypermultiplets
\cite{hyperm,baggerwitten}.
The new instantons that gauge-fix this theory
were called by us hyperinstantons and are the main subject
of the present paper.
Moreover, the twist
of N=2 supergravity coupled to hypermultiplets
gives a theory of topological gravity
coupled to topological $\sigma$-model in which the condition $\omega^{-ab}=0$
is modified by a
contribution due to the hypermultiplets. The hyperinstanton equations,
instead, are unmodified.

In this paper we formulate the topological $\sigma$-model in four
dimensions as a theory of maps $q:{\cal M}\rightarrow {\cal N}$
from a four dimensional Riemannian
manifold ${\cal M}$
to an almost quaternionic
manifold ${\cal N}$ (if ${\cal M}$ is Riemannian and four dimensional, then
it is also almost quaternionic \cite{salamon}).
Inspired by the results coming from the topological twist
\cite{anselmifre2,erice}, we propose the following concept
of triholomorphic maps $q:{\cal M}\rightarrow {\cal N}$. Let
$T{\cal M}$ and $T{\cal N}$ be the tangent bundles to ${\cal M}$ and
${\cal N}$.
Let $\{j_u:T{\cal M}\rightarrow T{\cal M}, u=1,2,3\}$ and
$\{J_u:T{\cal N}\rightarrow T{\cal N}, u=1,2,3\}$ be almost quaternionic
structures on ${\cal M}$ and ${\cal N}$, respectively,
namely triplets of $(1,1)$-tensors\footnotemark
\footnotetext{In general, these are {\sl locally}
defined $(1,1)$-tensors \cite{galicki}. For the moment we suppose that they
can be defined globally. We shall soon come back to this point.}
satisfying the quaternionic algebras
\begin{eqnarray}
j_u\circ j_v&=&-\delta_{uv}\,{\rm id}_{T{\cal M}}+\varepsilon_{uvz}j_z,
\nonumber\\
J_u\circ J_v&=&-\delta_{uv}\,{\rm id}_{T{\cal N}}+\varepsilon_{uvz}J_z,
\label{quaternionicalgebras}
\end{eqnarray}
where ${\rm id}_{T{\cal M}}$ and ${\rm id}_{T{\cal N}}$ are the identity
maps on $T{\cal M}$ and $T{\cal N}$, respectively. By convention,
$j_u$ (resp.\ $J_u$) will be called
the {\sl almost quaternionic $(1,1)$-tensors}
of ${\cal M}$ (resp.\ ${\cal N}$).

Consider the pull-forward $q_*:T{\cal M}\rightarrow
T{\cal N}$ of the map $q:{\cal M}\rightarrow {\cal N}$ and the
following diagram
\begin{equation}
\matrix{T{\cal M}&\stackrel{q_*}{\longrightarrow} &T{\cal N}\cr
${\it \scriptsize $j_u$}$\uparrow& &\downarrow ${\it \scriptsize $J_v$}$\cr
T{\cal M}&\stackrel{J_v\circ q_*\circ j_u}{\longrightarrow} &T{\cal N}}.
\end{equation}
We see that $J_v\circ q_*\circ j_u$ acts from $T{\cal M}$ to
$T{\cal N}$, precisely as $q_*$. Consider the equation
\begin{equation}
q_*-J_u\circ q_*\circ j_u=0.
\label{afeq}
\end{equation}
The sum over the repeated index $u$ is understood.
This is an equation on the map $q:{\cal M}
\rightarrow {\cal N}$ and it is our proposal
for the definition
of triholomorphic maps from a four dimensional Riemannian manifold
${\cal M}$ to an almost quaternionic manifold ${\cal N}$.
The definition (\ref{afeq})
of triholomorphic maps is not the only possible choice.
There is no uniqueness of the relative ordering
of the three almost quaternionic $(1,1)$-tensors of the two manifolds.
Moreover,
the almost quaternionic $(1,1)$-tensors are in general only locally defined
\cite{galicki}, i.e.\ defined on neighborhoods $U_{(\alpha)}$
such that on the intersection $U_{(\alpha)}\cap U_{(\beta)}$
of two neighborhoods the transition functions are $SO(3)$ matrices
$\Lambda^{uv}$.  Consequently, eq.\ ({\ref{afeq}) should be substituted by
the more general condition
\begin{equation}
q_*-\Lambda^{uv}J_u\circ q_*\circ j_v=0,
\label{afeq4}
\end{equation}
where $\Lambda$ is an $SO(3)$ matrix that can depend on the point.
Then,
triholomorphic maps are those maps $q$ such that there exists a $\Lambda^{uv}$
such that (\ref{afeq4}) holds.
We postpone the discussion of this ambiguity
to later sections, where we shall see
that it has a quite simple physical interpretation.

For eq.\ (\ref{afeq}) to be meaningful, it is
only required that both ${\cal M}$ and ${\cal N}$
possess almost
quaternionic structures.
The most interesting cases are when ${\cal M},{\cal N}$ are hyperK\"ahler
\cite{hyperk}
or quaternionic K\"ahler \cite{quaterni,galicki,dauriaferrarafre}
(the latter will be simply called
``quaternionic''), namely when the almost quaternionic structures possess
more properties.

We are going to show
that eq.\ (\ref{afeq}) is a good definition and
that it agrees with the concept of triholomorphic maps as
formulated in ref.s \cite{anselmifre2,erice} coming from the topological twist
(despite the explicit appearance of the metric in the equations that
naturally follow from the twist).

In section \ref{sigmamodel}
we formulate the general topological $\sigma$-model,
where the gauge-fixing is provided by the triholomorphic instanton
condition (\ref{afeq}). This topological $\sigma$-model
naturally describes the moduli-space of triholomorphic maps (that we
call hyperinstantons, when dealing with physics). In
section \ref{physics} we make the match with the
theories coming from the topological twist.
Indeed, we show that when ${\cal M}$ and ${\cal N}$ are both
hyperK\"ahler manifolds, the general topological $\sigma$-model
gauge-fixed by (\ref{afeq}) is nothing else but the topological
twist of the N=2 globally supersymmetric $\sigma$-model (hypermultiplets).
On the other hand, when either ${\cal M}$ or ${\cal N}$ is quaternionic
(external gravity), the topological $\sigma$-model gauge-fixed by
(\ref{afeq}) is not obtainable as the topological twist
of any N=2 globally supersymmetric theory. It is notorious
that a quaternionic ${\cal N}$ corresponds to the case of
a locally supersymmetric $\sigma$-model, so that the case
${\cal N}$ quaternionic requires dynamical gravity in the twisting
procedure. This simply shows the well-known fact
that the set of possible topological theories
is larger than the set of those obtainable from the twist.
The complete details of the
topological twist are shown in appendix \ref{twist}, while the
general definitions of hyperK\"ahler and quaternionic manifolds
can be found in appendix \ref{hyperappendix}.

As far as we know, eq.s (\ref{afeq})
have not been proposed in the mathematical literature, apart from
the case ${\cal M}={\cal N}={\bf R}^4$
where they reduce to the Cauchy-Fueter equations \cite{sudbery,gentili}.
In section \ref{trio} we recall the main properties of the solutions to the
Cauchy-Fueter equations, to give the reader an idea of what they are.
In section \ref{torus} we derive the general form
of triholomorphic maps
in the case ${\cal M}={\cal N}=T_4$.
In section \ref{k3} we exhibit a class of solutions with ${\cal M}
={\cal N} =K3$ (we consider the Fermat surface for simplicity)
to convince the reader that the set of solutions to these new equations
is non-empty and non-trivial. This class of solutions will be described
directly by the properties of the polynomial that defines the Fermat surface,
to address the possibility that eq.s (\ref{afeq}) have
an algebraic counterpart in the case of algebraic varieties.
One can also generalize the concept of ``rational curves" of a K\"ahler
manifold by choosing ${\cal M}=S^4$ or ${\bf CP}^2$ and studying the
triholomorphic
embeddings of these manifolds into hyperK\"ahler or quaternionic manifolds
of any dimension. Interesting embeddings are, of course, also those of
${\cal M}=T_4$ or K3.

Physics also provides a topological theory of dynamical gravity coupled
to the $\sigma$-model, as mentioned above, in which case the
target manifold ${\cal N}$ is quaternionic. The coupled equations (something
like the ``square root" of the coupled Einstein and matter equations)
are very difficult to solve, as one can expect. In section
\ref{dynamical} we show that the simplest ans\"atze are not solutions.
Nevertheless, the same ans\"atze
are instantons of the general
$\sigma$-model presented in the next section (with external gravity),
that also contemplates the case of a quaternionic ${\cal N}$.

When the world-manifold and the target one are both
hyperK\"ahler or quaternionic and four dimensional,
we can think of hyperinstantons as maps that go form a gravitational
instanton (the world-manifold) to a gravitational
instanton (the target) and that are themselves instantons.
So, the study of hyperinstantons may be useful for getting insight into
gravitational instantons (here intended as
manifolds with a self-dual Riemann tensor
or a self-dual Weyl tensor).
We hope that our work will stimulate research into this subject,
because we think that it can be source of insight into the problem of
gravitational instantons.

\section{Topological $\sigma$-model for triholomorphic maps}
\label{sigmamodel}

In this section we build the topological $\sigma$-model for hyperinstantons,
by generalizing the method used by Witten in ref.\ \cite{witten2}
and further clarified by Baulieu and Singer in ref.\ \cite{besi2}.
We need Riemannian metrics both
on ${\cal M}$ and ${\cal N}$. We suppose
that they are Hermitian with respect to the almost quaternionic $(1,1)$-tensors
of the corresponding manifolds.
We notice that any four dimensional Riemannian manifold can be naturally
endowed with an
almost quaternionic structure such that the metric is
Hermitian \cite{salamon}.

Introducing indices explicitly, equation (\ref{afeq})
takes the form
\begin{equation}
\partial_\mu q^i-{(j_u)_\mu}^\nu\partial_\nu q^j {(J_u)_j}^i=0,
\label{afeq2}
\end{equation}
where $\mu=1,\ldots 4$ are the world indices and $i=1,\ldots 4n$
are the target ones (we set ${\rm dim}\,{\cal N}=4n$).
We see that (\ref{afeq2}) is the natural generalization of the
Cauchy-Riemann equations (\ref{cauchyriemann}). The $16n$
equations (\ref{afeq2}) are not all independent. Indeed, we expect $4n$
equations. The correct counting is retrieved by observing that the
matrix
\begin{equation}
H^i_\mu=\partial_\mu q^i-{(j_u)_\mu}^\nu\partial_\nu q^j {(J_u)_j}^i
\end{equation}
satisfies identically the duality condition
\begin{equation}
H_\mu^i+{1\over 3}{(j_u)_\mu}^\nu H_\nu^j {(J_u)_j}^i=0.
\label{dualitycondition}
\end{equation}
This condition, as the reader can easily verify, reduces the
number of equations by a
factor four.

The BRST-quantization of the theory is achieved as follows. We introduce
topological ghosts $\xi^i$ (ghost number $g=1$), as well
as topological antighosts $\zeta_\mu^i$ ($g=-1$) and
Lagrange multipliers $b_\mu^i$ ($g=0$) for the gauge-fixing
(\ref{afeq2}). Antighosts and
Lagrange multipliers are required to satisfy the same duality condition
as the left hand side of (\ref{afeq2}),
namely
\begin{eqnarray}
\matrix{\zeta_\mu^i+{1\over 3}{(j_u)_\mu}^\nu \zeta_\nu^j {(J_u)_j}^i=0,&
b_\mu^i+{1\over 3}{(j_u)_\mu}^\nu b_\nu^j {(J_u)_j}^i=0.}
\label{selfduality}
\end{eqnarray}
The BRST operator will be denoted by $s$ and the BRST-variation
of $q^i$ will be $sq^i=\xi^i$, so that $s\xi^i=0$.
On the other hand, the BRST-variation
of the topological antighost $\zeta_\mu^i$ is not simply the
Lagrange multiplier $b_\mu^i$, since we have to make sure that the
duality condition of $\zeta_\mu^i$ is preserved by the BRST-algebra.
The correct form of $s\zeta_\mu^i$ is
$s\zeta_\mu^i=b_\mu^i-\Gamma^i_{jk}\xi^j\zeta_\mu^k
-{1\over 4}{(j_u)_\mu}^\nu {\cal D}_k {(J_u)_j}^i \xi^k\zeta_\nu^j$, where
$\Gamma^i_{jk}$ is the Levi-Civita connection on the target manifold
${\cal N}$, while ${\cal D}_k$ is the covariant derivative on ${\cal N}$.
The BRST-variation of $b_\mu^i$ is obtained by demanding
$s^2\zeta_\mu^i=0$. One then checks consistency by
verifying that $s^2 b_\mu^i=0$ and that the duality condition
(\ref{selfduality}) on $b_\mu^i$ is preserved. The complete BRST algebra is
given by
\begin{eqnarray}
sq^i&=&\xi^i,\nonumber\\
s\xi^i&=&0,\nonumber\\
s\zeta_\mu^i&=&b_\mu^i-\Gamma^i_{jk}\xi^j\zeta_\mu^k
-{1\over 4}{(j_u)_\mu}^\nu {\cal D}_k {(J_u)_j}^i \xi^k\zeta_\nu^j,
\nonumber\\
sb_\mu^i&=&{1\over 2}{{R_{jk}}^i}_l\xi^j\xi^k\zeta_\mu^l-
\Gamma^i_{jk}\xi^jb_\mu^k
-{1\over 4}{(j_u)_\mu}^\nu {\cal D}_k {(J_u)_j}^i \xi^kb_\nu^j
\nonumber\\&&
+{1\over 4}{(j_u)_\mu}^\nu {\cal D}_m{\cal D}_k {(J_u)_j}^i \xi^m\xi^k
\zeta_\nu^j
-{1\over 16} {\cal D}_k {(J_u)_j}^i{\cal D}_l {(J_u)_m}^j \xi^k\xi^l
\zeta_\mu^m\nonumber\\&&
+{1\over 16} \varepsilon_{uvz}{(j_z)_\mu}^\rho{\cal D}_k {(J_u)_j}^i
{\cal D}_l {(J_v)_m}^j \xi^k\xi^l\zeta_\rho^m.
\end{eqnarray}

To find a Lagrangian for the theory, we have to choose a gauge fermion
$\Psi$ that fixes the gauge according to eq.\ (\ref{afeq2}). This is
achieved by setting
\begin{equation}
\Psi=\int_{\cal M}d^4 x \, \sqrt{g}g^{\mu\nu}h_{ij}\zeta^i_\mu
\left(\partial_\nu q^j-{1\over 8}b_\nu^j\right).
\end{equation}
The action ${\cal S}=s\Psi$ then turns out to be
\begin{equation}
{\cal S}={\cal S}_{bosonic}+{\cal S}_{ghost},
\end{equation}
where
\begin{eqnarray}
{\cal S}_{bosonic}&=&\int_{\cal M}d^4x\, \sqrt{g}g^{\mu\nu}
h_{ij}b_\mu^i \left(\partial_\nu q^j-{1\over 8}b_\nu^j\right),
\nonumber\\
{\cal S}_{ghost}&=&\int_{\cal M}d^4x\, \sqrt{g}\left(
-g^{\mu\nu}h_{ij}\zeta^i_\mu{\cal D}_\nu\xi^j+{1\over 16}R_{ijkl}
g^{\mu\nu}\zeta^i_\mu
\zeta^j_\nu\xi^k\xi^l\right.\nonumber\\&&
+\left.
{1\over 4}\zeta^m_\rho (j_u)^{\nu\rho}{\cal D}_k(J_u)_{mj}\partial_\nu q^j
\xi^k+{1\over 32}\zeta^i_\mu\zeta^l_\rho
(j_u)^{\mu\rho}{\cal D}_m{\cal D}_k(J_u)_{li}\xi^m\xi^k\right.\nonumber\\&&
\left.
-{1\over 128}g^{\mu\nu}\zeta^i_\mu\zeta^m_\nu
{\cal D}_k(J_u)_{li}{\cal D}_n{(J_u)_m}^l\xi^k\xi^n
\right.\nonumber\\&&\left.
+{1\over 128}\zeta^i_\mu\zeta^m_\rho\varepsilon_{uvz}(j_z)^{\mu\rho}
{\cal D}_k(J_u)_{li}{\cal D}_n{(J_v)_m}^l\xi^k\xi^n\right).
\label{action}
\end{eqnarray}
The covariant derivative ${\cal D}_\mu\xi^i$ of $\xi^i$ is defined,
as usual, according to
\begin{equation}
{\cal D}_\mu\xi^i=\partial_\mu\xi^i+\Gamma^i_{jk}\partial_\mu q^j \xi^k.
\end{equation}

Keeping into account the selfduality condition (\ref{selfduality}),
the equation of motion of the Lagrange multiplier $b_\mu^i$ is
\begin{equation}
b_\mu^i=\partial_\mu q^i-{(j_u)_\mu}^\nu\partial_\nu q^j {(J_u)_j}^i.
\end{equation}
By eliminating the Lagrange multiplier from the action (\ref{action}),
one arrives at the following final form of the
bosonic action (the ghost action ${\cal S}_{ghost}$
is not affected),
\begin{eqnarray}
{\cal S}_{bosonic}&=&\int_{\cal M}d^4 x\, \left({1\over 2}\sqrt{g}
g^{\mu\nu}h_{ij}
\partial_\mu q^i\partial_\nu q^j+{1\over 2}\sqrt{g}
(j_u)^{\mu\nu}(J_u)_{ij}\partial_\mu q^i \partial_\nu q^j
\right).
\label{action2}
\end{eqnarray}
This is the usual $\sigma$-model action (\ref{saction}) plus the term
\begin{equation}
{\cal S}_T={1\over 2}\int_{\cal M}d^4 x\, \sqrt{g}
(j_u)^{\mu\nu}(J_u)_{ij}\partial_\mu q^i \partial_\nu q^j.
\label{stac}
\end{equation}
When ${\cal M}$ and ${\cal N}$
are both hyperK\"ahler manifolds, ${\cal S}_T$ is a topological invariant.
Indeed, in this case
let us introduce
the K\"ahler forms (see appendix \ref{hyperappendix})
\begin{eqnarray}
\Omega_u&=&{(J_u)_i}^jh_{jk}\, dq^i\wedge dq^k,\nonumber\\
\Theta_u&=&{(j_u)_\mu}^\nu g_{\nu\rho}\, dx^\mu\wedge dx^\rho.
\end{eqnarray}
$\Theta_u$ are selfdual or antiselfdual. We choose them to be
antiselfual ($^*\Theta_u=-\Theta_u$).
Then, (\ref{stac}) can be written as
\begin{equation}
{\cal S}_T=-{1\over 4}\int_{\cal M}q^*\Omega_u\wedge \Theta_u.
\label{stac0}
\end{equation}
The topological character of ${\cal S}_T$ is now evident, since
both $\Omega_u$ and $\Theta_u$ are closed.
The theory with ${\cal M}$ and ${\cal N}$
hyperK\"ahler is the theory that can be obtained
by topological twist (see section \ref{physics}).
In general, the quantum action ${\cal S}_q$ is the sum of a classical action
${\cal S}_{class}$ plus the BRST-variation $s\Psi$ of a gauge-fermion
$\Psi$. ${\cal S}_{class}$ should be a topological invariant. So far,
we have taken ${\cal S}_{class}=0$. In the case when ${\cal M}$ and ${\cal N}$
are hyperK\"ahler, a good classical action can be
${\cal S}_{class}=-{\cal S}_T$, so that the BRST-quantized action
${\cal S}=-{\cal S}_T+s\Psi$
is the usual $\sigma$-model action plus the ghost action.
However, this
cannot always be achieved, since in general ${\cal S}_T$
is not a topological invariant (e.g.\ when
either ${\cal M}$ or ${\cal N}$ are quaternionic).

The equation of the deformations $\delta q^i$ of the hyperinstanton condition
(\ref{afeq2}) is
\begin{eqnarray}
{\cal D}_\mu\delta q^i-{(j_u)_\mu}^\nu{\cal D}_\nu\delta q^j {(J_u)_j}^i
-{(j_u)_\mu}^\nu\partial_\nu q^j {\cal D}_k{(J_u)_j}^i\delta q^k
\nonumber\\
-\Gamma^i_{jk}(\partial_\mu q^j-{(j_u)_\mu}^\nu\partial_\nu q^l
{(J_u)_l}^j)\delta q^k=0,
\end{eqnarray}
Consequently, a deformation $q^i+\delta q^i$ of a solution $q^i$
to eq.\ (\ref{afeq}) still satisfies the triholomorphicity
condition (\ref{afeq}) if and only if
\begin{equation}
D_\mu\delta q^i\equiv
{\cal D}_\mu\delta q^i-{(j_u)_\mu}^\nu{\cal D}_\nu\delta q^j {(J_u)_j}^i
-{(j_u)_\mu}^\nu\partial_\nu q^j {\cal D}_k{(J_u)_j}^i\delta q^k=0.
\label{zeromodes}
\end{equation}
This is also the equation of the zero modes of the topological ghosts
$\xi^i$, namely their field equation, linearized in the Fermi fields
and calculated in the instantonic background.
The ghost number anomaly $\Delta g$ is the index of the operator $D_\mu$
and is also called the formal dimension of the moduli-space.

Let us discuss the observables of the theory.
Any closed and nonexact form $\Omega$ on the target manifold
generates (after pull-back to the world-manifold)
descent equations and nontrivial observables, {\sl via} the BRST-extension
of the identity
$d\, \Omega=0$. Let $\Omega_p^{\alpha_p}$ be
a representative of a $p$-cocycle on the target manifold ${\cal N}$,
$0\leq p\leq 4n={\rm dim}\, {\cal N}$, $1\leq{\alpha_p}\leq b^p({\cal N})$,
where $b^p({\cal N})$ denote the Betti numbers of ${\cal N}$.
Let $\hat \Omega_p^{\alpha_p}$ be the BRST-extension
of $\Omega_p^{\alpha_p}$, namely
\begin{equation}
\hat \Omega_p^{\alpha_p}=\Omega^{\alpha_p}_{(p,0)}+\Omega^{\alpha_p}_{(p-1,1)}
+\ldots +\Omega^{\alpha_p}_{(0,p)},
\end{equation}
where $\Omega^{\alpha_p}_{(p,0)}=\Omega^{\alpha_p}_p$ and the terms
$\Omega^{\alpha_p}_{(p-k,k)}$ are obtained by substituting $k$ differentials
$dq^i$ with the topological ghosts $\xi^i=s q^i$.
Let $\hat d$ be the sum of the exterior derivative operator $d$ plus
the BRST operator $s$, $\hat d=d+s$.
The descent equations
$\hat d \hat \Omega^{\alpha_p}_p=0$ read
\begin{equation}
d\Omega^{\alpha_p}_{(p-k,k)}+s\Omega^{\alpha_p}_{(p-k+1,k-1)}=0,
\end{equation}
for $k=0,\ldots p+1$ and with the convention
$\Omega^{\alpha_p}_{(p+1,-1)}=
\Omega^{\alpha_p}_{(-1,p+1)}=0$. Thus we have the following observables
\begin{equation}
{\cal O}^{\alpha_p,\beta^{k}}_{p,k}\equiv\int_{\gamma^{\beta^{k}}_{k}}
q^*\Omega^{\alpha_p}_{(k,p-k)},
\end{equation}
where $\gamma^{\beta^{k}}_{k}$ is a basis of $k$-cycles
on the world-manifold ${\cal M}$,
$\beta^{k}=1,\ldots b_{k}({\cal M})$.
Notice that the property $q^*\circ d=d\circ q^*$ is BRST extended
to $q^*\circ \hat d=\hat d\circ q^*$ and so we also have
$q^*s=sq^*$.

The physical amplitudes of the theory are average values of products
of observables
\begin{equation}
<{\cal O}_1\cdots {\cal O}_l>.
\end{equation}
If $g_i$, $i=1,\ldots l$ is the ghost number of ${\cal O}_i$, the condition
for the amplitude to be possibly nonvanishing
is that the sum $\sum_{i=1}^l g_i$ must be equal to the ghost number
anomaly $\Delta g$.
The physical amplitudes are topological invariants of
the moduli-space. They generalize the
Donaldson polynomials that one finds in pure topological
Yang-Mills theory \cite{witten,donaldson}.

A topological theory can be deformed by adding extra terms to the
action. We call {\sl deformation} any BRST-invariant term appearing
in the quantum action, in addition to the BRST-exact term $s\Psi$.
It must be a BRST-invariant integral of a
world-four-form over ${\cal M}$.  Let $\omega_k^{\beta_k}$ denote
a basis of
world-cocycles of degree $k=1,\ldots 4$, $\beta_k=1,\ldots b^k({\cal M})$.
Then the most general deformation is a linear combination of
\begin{equation}
{\cal D}^{\alpha_p,\beta_{k}}_{p,k}\equiv
\int_{\cal M}q^*\Omega^{\alpha_p}_{(4-k,p-4+k)}\wedge \omega^{\beta_k}_k.
\label{mostgen}
\end{equation}
So, the partition function
\begin{equation}
Z=\int d\mu \,{\rm exp}\left\{-s\Psi\right\}
\end{equation}
can be deformed into
\begin{equation}
Z=\int d\mu \,{\rm exp}\left\{-s\Psi+
\sum_{k=0}^4\sum_{p=4-k}^{4n}\sum_{\beta_k=1}^{b^k({\cal M})}
\sum_{\alpha_p=1}^{b^p({\cal N})}
s^{p,k}_{\alpha_p,\beta_k}{\cal D}^{\alpha_p,\beta_k}_{p,k}\right\},
\label{defsimod}
\end{equation}
where $s^{p,k}_{\alpha_p,\beta_k}$ are
the parameters of the deformations. Similarly, the deformed amplitudes
are
\begin{equation}
<{\cal O}_1\cdots {\cal O}_k \, {\rm exp}\left\{
\sum_{k=0}^4\sum_{p=4-k}^{4n}\sum_{\beta_k=1}^{b^k({\cal M})}
\sum_{\alpha_{p}=1}^{b^p({\cal N})}
s^{p,k}_{\alpha_p,\beta_k}{\cal D}^{\alpha_p,\beta_k}_{p,k}\right\}>.
\end{equation}
If ${\cal M}$ is compact, the Poincar\'e duality theorem says that
there exists a world-$(4-k)$-cycle, say $\gamma^{\beta_k}_{4-k}$, such that
\begin{equation}
{\cal D}^{\alpha_p,\beta_k}_{p,k}=\int_{\gamma^{\beta_k}_{4-k}}
q^*\Omega^{\alpha_p}_{(4-k,p-4+k)}=
{\cal O}_{p,4-k}^{\alpha_p,\beta_k}.
\end{equation}
The total number $d$ of deformations ${\cal D}^{\alpha_p,\beta_k}_{p,k}$
(which is the same as the total number
of observables, when ${\cal M}$ is compact) is
\begin{equation}
d=\sum_{q=0}^4 b^q({\cal M})\sum_{p=q}^{4n}b^p({\cal N}).
\end{equation}
In general, the total number of observables ${\cal O}_{p,k}^{\alpha_p,\beta^k}$
is given by a similar formula, where $b^q({\cal M})$ is replaced
by $b_q({\cal M})$.

\section{Relation with N=2 theories through the topological twist}
\label{physics}

In this section we discuss the topological field theories
that are originated by twisting the N=2 supersymmetric $\sigma$-models
(both the case of external and dynamical gravity).
We show that in the case of external gravity the twisted topological
$\sigma$-model corresponds to the general $\sigma$-model
that was formulated in the previous section, provided
${\cal M}$ and ${\cal N}$ are both hyperK\"ahler. When
either ${\cal M}$ or ${\cal N}$ are quaternionic and
gravity is external, we have not an N=2 theory generating the
corresponding topological $\sigma$-model by twist. Indeed, a
quaternionic ${\cal N}$ requires coupling to supergravity in order
to exhibit N=2 supersymmetry. The equations that one obtains by twisting
N=2 supergravity coupled to hypermultiplets
are more general than (\ref{afeq}), since they
are also required to gauge-fix the world-metric.
For a discussion of these equations, see section \ref{dynamical}.

Let $V^a=V^a_\mu dx^\mu$ be a vierbein for
the world-manifold ${\cal M}$ ($g_{\mu\nu}=V^a_\mu V_{\nu a}$)
and $E^{ak}=E^{ak}_i dq^i$ be a vielbein
for the target-manifold
${\cal N}$ ($h_{ij}=2E^{ak}_iE_{j\, ak}$ in our notation).
This way of writing the target vielbein ($k=1,\ldots n$ if
${\rm dim}\,{\cal N}=4n$) in which the index $a$ is identified with the
indices of the Lorentz group of the world-manifold comes naturally from the
twist \cite{anselmifre2}, as we show in detail in appendix
\ref{twist} and can always be done, at least locally.

Let us start by analysing the theory that comes from the N=2
globally supersymmetric $\sigma$-model (hypermultiplets). In this case
${\cal N}$ is hyperK\"ahler and ${\cal M}$ should be flat; however,
we use a covariantized notation, since we expect
the topological theory to be defined on more general world-manifolds.
In fact, as anticipated, the topological theory is well-defined for any
hyperK\"ahler world-manifold ${\cal M}$. We choose a Lorentz frame
for ${\cal M}$ such that $\omega^{-ab}=0$.
The hyperinstanton equations \cite{anselmifre2} read
\begin{eqnarray}
\matrix{V^{\mu[a}E_i^{b]^+k}\partial_\mu q^i =0,&
V^\mu_a E^{ak}_i \partial_\mu q^i =0.}
\label{inst1}
\end{eqnarray}
$[ab]^+$ means antisymmetrization and seldualization in the indices
$a,b$.

The proof that hyperinstantons minimize the action consists
in showing that the total action is a sum of squares of the left hand
sides of equations (\ref{inst1}) plus a total derivative \cite{anselmifre2},
namely\footnotemark\footnotetext{We retain the Minkowskian notation,
namely the notation that comes naturally from the starting N=2 theory.
In the formul\ae\ that come directly from the topological twist,
the Wick rotation to the Euclidean region will be understood.
The parameter $\lambda$ also comes from the formulation
of N=2 supergravity coupled to hypermultiplets, as elaborated in ref.\
\cite{dauriaferrarafre}; see also the appendices. We retain it
to facilitate the comparison between the various formul\ae.}
\begin{eqnarray}
\int_{\cal M} &d^4x\, 4\lambda
\sqrt{-g}& g^{\mu\nu}\partial_\mu q^i\partial_\nu q^jh_{ij}=
8\lambda\int_{\cal M} d^4x
\sqrt{-g}[(V^\mu_a E^{ak}_i \partial_\mu q^i)^2+4(V^{\mu[a}E_i^{b]^+k}
\partial_\mu q^i)^2]\nonumber\\
&+16i\lambda\int_{\cal M}&E^{[a k}\wedge E^{b]^- k}\wedge V_a \wedge V_b.
\label{dim1}
\end{eqnarray}
$E^{[a k}\wedge E^{b]^- k}$ is proportional
to the contraction of the Pauli matrices $\sigma_u$ with
the K\"ahler forms $\Omega_u$ of the target manifold ${\cal N}$
(see formula (\ref{kalviel}) of appendix \ref{hyperappendix}). Consequently,
$E^{[a k}\wedge E^{b]^- k}\wedge V_a \wedge V_b$ is a linear combination of
$\Omega^u\wedge V_{[a} \wedge V_{b]^-}$.
$\Omega_u$ are closed forms if ${\cal N}$ is
hyperK\"ahler. Using this fact one can easily
show that the last term in
(\ref{dim1}) is a topological invariant, provided
$\omega^{-ab}=0$. This is the reason why ${\cal M}$ cannot be any
four dimensional manifold, but it is required to
be hyperK\"ahler\footnotemark\footnotetext{The fact that the one has to be
cautios
when covariantizing the theory obtained by twisting the
N=2 supersymmetric $\sigma$-model
has been recently
confirmed by the results of ref.\ \cite{labastida}.
We recall, on the other hand, that no similar problem arises in the case
of topological Yang-Mills theory \cite{witten}.}.
We conclude that in case of external gravity
the action of the topological theory is the sum of the classical action
\begin{equation}
{\cal S}_T=16i\lambda\int_{\cal M}
E^{[a k}\wedge E^{b]^- k}\wedge V_a \wedge V_b,
\label{st0}
\end{equation}
plus the squares of the gauge-fixings plus the ghost terms,
i.e.\
\begin{equation}
{\cal S}={\cal S}_T+s\Psi.
\end{equation}
We can thus distinguish the following two cases.

i) ${\cal M}$ is not hyperK\"ahler. Then ${\cal S}_T$ is not a topological
invariant and we are not guaranteed that the solutions to (\ref{inst1})
solve the field equations, so we cannot consider (\ref{inst1}) as good
instantonic equations.
${\cal S}_T$ depends on the world-metric and thus the energy-momentum
tensor $T_{\mu\nu}$ is not BRST-exact\footnotemark
\footnotetext{In ref.\ \cite{labastida} an analysis of the breaking of
topological symmetry due to the non-BRST-exactness of $T_{\mu\nu}$ can be
found.}.
We are lead to conclude that the twisted-covariantized theory
is not a good topological field theory.

ii) ${\cal M}$ is hyperK\"ahler. Then ${\cal S}_T$ is a topological invariant
and (\ref{inst1}) are good instantonic equations. The twisted theory
is a well-defined topological $\sigma$-model.
We shall prove in a moment that
${\cal S}_T$
corresponds to (\ref{stac0}), which
is clearly independent of the ${\cal M}$ metric.
In any case, one can always get rid of ${\cal S}_T$ by
deciding that the quantum action is not ${\cal S}_T+s\Psi$, but simply
$s\Psi$, as in the previous section.
The difference, being a topological invariant, is
immaterial from the point of view of the N=2 theory. This change of action
can be also viewed as a deformation of the topological theory
of the kind ${\cal D}_{2,2}^{\alpha_2,\beta_2}$.

We now show that eq.s (\ref{inst1}) are equivalent to eq.s
(\ref{afeq}). Let us introduce
three matrices $I_u^{ab}$ that are antiselfdual in $ab$ and
satisfy the quaternionic algebra
\begin{equation}
I_uI_v=-\delta_{uv}+\varepsilon_{uvz}I_z.
\label{complexstructures}
\end{equation}
For future use, we fix an explicit form for these matrices, for example
\begin{equation}
\matrix{I_1=\left(\matrix{0&1&0&0\cr -1&0&0&0\cr
0&0&0&-1\cr 0&0&1&0}\right),&
I_2=\left(\matrix{0&0&1&0\cr 0&0&0&1\cr -1&0&0&0\cr
0&-1&0&0}\right),&
I_3=\left(\matrix{0&0&0&1\cr 0&0&-1&0\cr  0&1&0&0\cr
-1&0&0&0}\right).}
\label{opuwe}
\end{equation}
Let
\begin{equation}
A^{ab}_k\equiv V^{\mu a}E^{bk}_i\partial_\mu q^i.
\end{equation}
Equations (\ref{inst1}) can be written as
\begin{equation}
\matrix{{A_k^a}_{\, a}=0,& A_k^{[ab]^+}=0.}
\end{equation}
These are $4n$ equations and can be grouped together into
\begin{equation}
A_k-I_uA_kI_u=0,
\end{equation}
which are indeed $4n$ independent equations,  because of a duality condition
similar to the one in eq.\ (\ref{dualitycondition}).
Then we can write
\begin{equation}
(A_k-I_uA_kI_u)^{ab}=V^{\mu a}E^{bk}_i
(\partial_\mu q^i-{(j_u)_\mu}^\nu\partial_\nu q^j {(J_u)_j}^i),
\end{equation}
where ${(j_u)_\mu}^\nu=I_u^{ab}V_{\mu a}V^\nu_b$ and
${(J_u)_j}^i={(I_u)_a}^bE^{ak}_j E^i_{bk}$ are the three almost
quaternionic $(1,1)$-tensors
of ${\cal M}$ and ${\cal N}$, respectively,
(compare with formula (\ref{kalviel}) for the proof).
Thus we have shown that the hyperinstanton
equations (\ref{inst1}) are equivalent to the triholomorphicity condition
(\ref{afeq}). At this point, it is simple to check that (\ref{st0})
corresponds to (\ref{stac0}).

In the case when gravity is dynamical, there is a further equation that adds
to (\ref{inst1}) and defines the gravitational
instantons of the theory, namely the condition
on the world metric, that turns out to be \cite{anselmifre2}
\begin{equation}
\omega^{-ab}+{1\over 2}I_u^{ab}q^*\omega^u=0.
\label{inst2}
\end{equation}
Here $\omega^u$, $u=1,2,3$ denote the $Sp(1)$ connection of the quaternionic
target manifold ${\cal N}$, $\Omega^u$ being the corresponding field strength
(see appendix \ref{hyperappendix}).
Equation (\ref{inst2}) implies
\begin{equation}
R^{-ab}=-{1\over 2}I_u^{ab}q^*\Omega^u,
\end{equation}
which is the generalization of the self-duality condition on the
Riemann tensor. If the target manifold is four dimensional, we can
also write $\omega^{-ab}=\tilde \omega^{-ab}$ or
$R^{-ab}=\tilde R^{-ab}$, where the tilded forms are the pull-backs of the
corresponding target forms. So, in the case of four dimensional
target manifold, the first equation of (\ref{inst1}) simply states that
the anti-self-dual part of the world-manifold spin connection
is equal to the pull-back of the anti-self-dual part of the target
spin-connection.
The analogous statement on the anti-self-dual part of the Riemann tensor
will be useful in section \ref{dynamical}. Equations (\ref{inst1})
can still be rewritten in the form (\ref{afeq2}), as before.

The total kinetic lagrangian (Einstein lagrangian plus $\sigma$-model
kinetic lagrangian)
can be written as a sum of squares of the left hand sides of
equations (\ref{inst1}) and (\ref{inst2}) plus a total derivative, namely
\begin{eqnarray}
{\cal L}_{kin}&=&\varepsilon_{abcd}R^{ab}\wedge V^c\wedge V^d
-{1\over 6}\lambda\varepsilon_{abcd}V^a\wedge V^b\wedge V^c\wedge V^d
g^{\mu\nu}h_{ij}\partial_\mu q^i\partial_\nu q^j=\nonumber\\
&=&4i\left(\omega^{-ab}+{1\over 2}I^{ab}_u q^*\omega^u\right)\wedge
\left(\omega_{-ac}+{1\over 2}({I_v})_{ac} q^*\omega^v\right)
\wedge V_b\wedge V^c+
\nonumber\\
&-&{\lambda\over 3}\varepsilon_{cdef}V^c\wedge V^d\wedge V^e\wedge V^f
[4(V^{\mu[a}E_i^{b]^+k}\partial_\mu q^i)^2+
(V^\mu_a E^{ak}_i \partial_\mu q^i)^2]+\nonumber\\
&+&\int_{\cal M} d[\varepsilon_{abcd}\omega^{ab}\wedge V^c
\wedge V^d
+2iV^a\wedge dV^a-2 iI_u^{ab} q^*\omega^u\wedge V_a\wedge V_b].
\end{eqnarray}
The last term represents the topological action, which is also expressed by
\begin{equation}
{\cal S}_T=-4i \int_{\cal M} d\left[\left(\omega^{-ab}+{1\over 2}
I_u^{ab}q^*\omega^u\right)\wedge V_a \wedge V_b\right].
\label{st1}
\end{equation}
We see that this expression is zero for any hyperinstanton, due
to (\ref{inst2}). The coupled action is indeed zero on any solution to the
field
equations. In this sense, gravitational instantons are not privileged with
respect to any solution to the field equations (differently from what
happens for Yang-Mills instantons). Only in the topological version
of the theory they are privileged, because of the topological gauge-fixing.

The case of external gravity can be considered as a particular case
of  quaternionic ${\cal N}$.
To perform the limit from hyperK\"ahler to quaternionic
manifold, one substitutes $\Omega_u$ with $\lambda \Omega_u$
everywhere, simplifies the $\lambda$'s
in all the formul\ae\ in which it is possible and then
lets $\lambda$ go to zero. $\omega^u$
are set to zero ($\Omega^u$ become closed) and so
equation (\ref{inst2}) reduces to $\omega^{-ab}=0$ for the world-manifold.
The almost quaternionic $(1,1)$-tensors become (globally defined)
covariantly constant complex structures.

Let us discuss the observables of the topological theories under
consideration. When gravity is dynamical, in addition to the observables of the
topological $\sigma$-model that we have exhibited
at the end of the previous section, there are observables
due to topological gravity.
They are generated
as usual by the identities $\hat d \,{\rm tr}
[\hat R\wedge \hat R]=0$
and $\hat d\, {\rm tr}[\hat R\wedge \hat {\tilde R}]=0$ \cite{anselmifre}
($R$ denotes the world-Riemann tensor and the trace
refers to the Lorentz indices).
In the case when
topological Yang-Mills theory is also coupled \cite{anselmifre2,erice},
there are also
observables generated by identities like $\hat d \,{\rm tr}
[\hat F\wedge \hat F]=0$, $\hat d \,{\rm tr}
[\hat F\wedge \hat F\wedge \hat F \wedge \hat F]=0$, and so on
(the total number of possibilities being the
rank of the gauge group), where $F$ is the field strength of the
matter vectors and the trace refers to gauge indices.

The condition for the average value $<{\cal O}_1\cdots {\cal O}_k>$
of observables ${\cal O}_i$
with ghost numbers $g_i$ to be non-vanishing is
$\sum_{i=1}^kg_i=\Delta g$, where $\Delta g$ is the ghost number anomaly.
In the theories coming from the topological twist, $\Delta g$
can also be viewed as the R-duality anomaly of the starting N=2 theory,
since it is R-duality that defines the ghost number of the twisted theory
\cite{anselmifre2}. So far, a complete analysis of this anomaly
has not appeared in the literature.

Notice that the physical amplitudes of the theory of topological
gravity (coupled or not to topological Yang-Mills and topological
$\sigma$-models)
that one gets by twisting N=2 supergravity
\cite{anselmifre,bardonecchia,anselmifre2,erice},
correspond to well-defined (because topological)
amplitudes of a nonrenormalizable theory (N=2 supergravity). So,
even if quantum gravity does not still exist, there is a subset
of physical amplitudes that are well-defined and calculable.
This interesting property is not common either
to the theories of topological gravity
that are formulated ``by hand'', i.e.\ without any topological twist
\cite{topG4}, or to any other topological field theory
obtained by twist  \cite{N=2tw}
(in general they are twisted versions of renormalizable
field theories). That is why we think that our theory of
topological gravity deserves
particular investigation.

Recalling (\ref{defsimod}), we have that in the most general case,
i.e.\ topological gravity coupled to topological Yang-Mills theory and to
topological sigma model, the partition function
can be deformed by adding ${\cal D}^{\alpha_p,\beta_k}_{p,k}$
(\ref{mostgen}) or the
other deformations that can be built in a similar way
from the descent equations
generated by the identities $\hat d \,{\rm tr}
[\hat R\wedge \hat R]=0$,
$\hat d\, {\rm tr}[\hat R\wedge \hat {\tilde R}]=0$, $\hat d \,{\rm tr}
[\hat F\wedge \hat F]=0$,
$\hat d \,{\rm tr}
[\hat F\wedge \hat F\wedge \hat F\wedge \hat F]=0$  and so on.

We conclude this section with some remarks about equations (\ref{inst1})
and (\ref{inst2}).
The hyperinstanton equations (\ref{inst1}) are invariant under
diffeomorphisms of the world-manifold
${\cal M}$.
However, they break local
Lorentz invariance. Moreover, equations (\ref{inst2}) are a generalization
of the equations of gravitational instantons $\omega^{-ab}=0$,
which also break local Lorentz
invariance. The deep reason of the breakdown becomes
clearer when considered
from the point of view of the topogical twist, as described
in appendix \ref{twist} and ref.s \cite{anselmifre,anselmifre2,erice}.
When gravity is dynamical, the redefinition of the Lorentz group
according to the rules
of the twist, mixes up the
original local Lorentz symmetry  with some global $SU(2)_Q\otimes SU(2)_I$.
This explains the loss of local invariance under the new Lorentz group.
Moreover, there is an arbitrariness which is intrinsic to
the definition of the twist, consisting in the choice of the reference frame
for $SU(2)_Q\otimes SU(2)_I$ before the twisting identifications.
This freedom will be called {\sl the relative Lorentz gauge} of
${\cal N}$ with respect to ${\cal M}$.

The hyperinstanton problems coming from the twist
can thus be stated as follows.

i) External gravity: given a four dimensional hyperK\"ahler manifold
${\cal M}$, with a metric and a Lorentz frame such that $\omega^{-ab}=0$,
and a hyperK\"ahler manifold ${\cal N}$, one has to find a map
$q:{\cal M}\rightarrow {\cal N}$ such that there exists a relative Lorentz
gauge of ${\cal M}$ with respect to ${\cal N}$ such that (\ref{inst1})
are satisfied;

ii) dynamical gravity: given a four dimensional topological space
${\cal M}$ and a
quaternionic manifold ${\cal N}$, one has to find
a metric and a Lorentz frame for
${\cal M}$ and a map $q:{\cal M}\rightarrow {\cal N}$ such that
there exist a relative Lorentz gauge of ${\cal M}$ with respect
to ${\cal N}$ such that
equations (\ref{inst1}) and (\ref{inst2}) are satisfied.

However, it can happen that different solutions to the hyperinstanton problem
do not contribute to the same topological field theory.
It is when two different solutions require different relative Lorentz gauges.
Indeed,
the relative Lorentz gauge is a fundamental point of the
topological theory.
It must be fixed once for all when defining the theory, because it enters
in the expression of the topological gauge-fixing.
Changing the relative Lorentz gauge
is equivalent to changing the topological theory.
We do not know, so far,
whether this is a meaningful change of theory or not.
In all the examples discussed in this paper it turns
out that this is not an essential change. The properties of the relative
Lorentz
gauge are well illustrated in the examples of section \ref{k3}.

In the topological $\sigma$-model that we formulated
in section \ref{sigmamodel}, we never introduced Lorentz indices. The
concept of relative Lorentz gauge is substituted by
the ambiguity $\Lambda^{uv}$ of equations (\ref{afeq})-(\ref{afeq4}),
since there is no preferred way to contract the index $u$ of the
world complex structures with the same index of the target ones.
Eq.\ (\ref{afeq}) is the simplest choice, but in the following
it will be convenient to deal with a more general choice of the
kind (\ref{afeq4}). The relative Lorentz gauge is thus the
physical interpretation of the intrinsic ambiguity of the
triholomorphicity conditions.

Since it is possible to formulate the topological $\sigma$-model with the
method of section \ref{sigmamodel}, at least in the case of external gravity
(so far, we do not possess a twist-independent formulation of
topological $\sigma$-models coupled to topological gravity), it is evident
that the breakdown of local Lorentz invariance is not
a meaningful breakdown.
Moreover, it is also clear that the splitting of the Lorentz index
of the target vielbein $E_i^{ak}$ into a couple of indices, one of which
is identified with the Lorentz indices of the world-manifold,
can be simply considered
as a convenient intermediate step, but does not put restrictions
on the target manifold.

\section{Triholomorphic functions}
\label{trio}

In this section we study the case in which both the world-manifold and
the target space are ${\bf R}^4\approx {\bf H}$. Moreover, at the
end of the section we prove
very simple general theorems about the solutions to (\ref{afeq})
 that generalize analogous theorems about holomorphic functions.

The elements $x\in {\bf H}$ are called {\sl quaternionic numbers}.
Reverting to
Euclidean signature, equations (\ref{inst1}) reduce to
\begin{equation}
\matrix{\partial_\mu q_\mu=0,&
\partial_\mu q_\nu-\partial_\nu q_\mu+\varepsilon_{\mu\nu\rho\sigma}
\partial_\rho q_\sigma=0.}
\label{flat}
\end{equation}
At first sight, one is tempted to interpret these equations as the
anti-self-duality condition on the field strength of some four-potential
$q_\mu$ in the Lorentz gauge. However, $q_\mu$ has nothing to do
with a four-potential and the first of Eq.\ (\ref{flat}) is not a gauge choice,
but a true equation. In fact, the correct interpretation of Eq.s
(\ref{flat}) is quite different, since
these equations are equivalent to the
Cauchy-Fueter equations
that define holomorphicity in the quaternionic sense and
that generalize the Cauchy-Riemann equations.
Let $I$, $J$ and $K$ be a representation of the quaternionic algebra,
\begin{equation}
\matrix{I^2=J^2=K^2=-1,\,\,\,&
IJ=-JI=K \,\,\,\,\,\, \&\,\, {\rm cyclic} \,\, {\rm permutations}.}
\label{complexsrtuctures}
\end{equation}
Let
\begin{equation}
\matrix{q=I q_1+J q_2+ K q_3+q_4,&
\bar\partial=-I\partial_1-J\partial_2-K \partial_3+\partial_4,}
\end{equation}
where $\partial_\mu=\partial/\partial x^\mu$.
Then equations (\ref{flat}) can be written in the form
\begin{equation}
\bar \partial q=0
\label{cauchyfueter}
\end{equation}
which are indeed the Cauchy-Fueter equations \cite{sudbery,gentili}.

A common representation of the complex structures (\ref{complex})
is given by
\begin{equation}
\matrix{I=-i\sigma_1&J=-i\sigma_2&K=-i\sigma_3}
\end{equation}
where $\sigma_1,\sigma_2,\sigma_3$ are the Pauli matrices. If $e^\mu=
(I,J,K,1)$ and $\bar e^\mu=(-I,-J,-K,1)$, then we have
$q=q_\mu e^\mu$ and $\bar\partial=\bar e^\mu\partial_\mu$.

The Cauchy-Fueter equations have not a unique form, due
to the ambiguity noticed in eq.\ (\ref{afeq4}). For
example, an alternative choice is
\begin{equation}
(-i\partial_1+j\partial_2+k\partial_3+\partial_4)q=0.
\label{flat1}
\end{equation}
It is clear that the identity $q=x=Ix_1+Jx_2+Kx_3+x_4$
is a solution of the new equations (\ref{flat1}), even if
it is not a solution to the old ones (\ref{cauchyfueter}). It is also
clear that essentially no {\sl new} solution is {\sl created}
by this trick. The set of solutions to (\ref{flat}) and (\ref{flat1}) are into
one-to-one correspondence.
The non-uniqueness of the form of the Cauchy-Fueter equations, namely the
lack of a canonical form among the possible ones such
as (\ref{cauchyfueter}), (\ref{flat1}) and similar,
has a very simple interpretation in our description, as we
remarked at the end of the previous section, namely
it is the relative Lorentz gauge of the world-manifold with respect
to the target one.

The operator $\bar\partial$ can be thought as the Weyl operator in the
Euclidean signature.
Indeed, the chiral Euclidean representation of the Dirac matrices
is
\begin{equation}
\matrix{
\gamma_i=\left(\matrix{0&i\sigma_i\cr -i\sigma_i&0}\right),&
\gamma_4=\left(\matrix{0&1\cr 1&0}\right),}
\end{equation}
with $i=1,2,3$, so that the Dirac operator is
\begin{equation}
\partial\!\!\!\slash=\left(\matrix{0&\bar \partial\cr \partial & 0}\right),
\end{equation}
where $\partial=e^\mu\partial_\mu$.
The Weyl equation for right handed Weyl spinors is simply
$\bar \partial\psi=0$. The Weyl spinor
$\psi$ is a doublet of complex numbers and can be interpreted as a
quaternionic number $q$
via the isomorphism ${\bf H}\approx{\bf C}\oplus J{\bf C}
=({\bf R}\oplus I {\bf R})\oplus J ({\bf R}\oplus I {\bf R})$.
In this sense the Cauchy-Fueter equations $\bar\partial q=0$ correspond
to the Weyl equation $\bar\partial \psi=0$.

This remark extends to the case in which
the target manifold ${\cal N}$ is ${\bf H}$
and the world-manifold ${\cal M}$ is a four dimensional
generic hyperK\"ahler manifold (we choose it to satisfy $\omega^{-ab}=0$).
Then the hyperinstanton equations are still equivalent to the Dirac equation
${\cal D}\!\!\!\!\slash \,\psi=0$,
on a right handed Weyl spinor $\psi=\left[0,0,\psi_1, \psi_2
\right]$, $\psi_1,\psi_2\in {\bf C}$ that
parametrizes ${\bf C}\oplus J{\bf C}\approx {\bf H}$. Noticing that on such
a spinor the covariant derivative is the same as the simple
exterior derivative (since $\omega^{-ab}=0$) and defining
\begin{equation}
\matrix{
q_1=-{\rm Im}\,\psi_2,&
q_2={\rm Re}\,\psi_2,\cr
q_3=-{\rm Im}\,\psi_1,&
q_4={\rm Re}\,\psi_1,}
\label{uij}
\end{equation}
we see that ${\cal D}\!\!\!\!\slash\, \psi=0$ becomes
\begin{equation}
\matrix{V^\mu_a\partial_\mu q^a=0,&
V^{\mu [a}\partial_\mu q^{b]^+}=0,}
\end{equation}
as expected. On the other hand, the correspondence between the
Dirac equation
for right handed Weyl spinors and the hyperinstanton equations is quite natural
in the case ${\cal N}$ is flat.
Indeed,
since the hyperinstanton equations are linear in $q^i$, they
have exactly the same form as the
equations of their deformations $\delta q^i$. Consequently,
they also have the same form as the equation of the zero modes $\zeta^a$
of the topological ghosts. On the other hand, in the twisting procedure,
the topological ghosts originate
from fermions $\zeta^I$, and the equation
of their zero modes comes from the Dirac equations ${\cal D}\!\!\!\!\slash
\,\zeta^I=0$. Hence, it is no wonder
that the Cauchy-Fueter equations correspond to the Dirac equation
in the case ${\cal N}$ is flat. To be precise,
the Dirac equations ${\cal D}\!\!\!\!\slash
\,\zeta^I=0$ can be put into the form
${\cal D}\!\!\!\!\slash\,\psi=0$,
if the four real components
of the right handed Weyl spinor $\psi$
are regarded as the four right handed
components of two Majorana spinors $\zeta^I$, $I=1,2$, via
the formula $(\zeta^{\dot \alpha})^I=
(\bar \sigma_a)^{\dot \alpha I}q^a$, $q^a$ being related to
$\psi_1$ and $\psi_2$ by (\ref{uij}). We then see that
${\cal D}\!\!\!\!\slash \,\psi=0$ is equivalent to
${\cal D}\!\!\!\!\slash \,\zeta^I=0$ and thus to the hyperinstanton equations,
as claimed. This interpretation of the hyperinstanton equations, however,
cannot be extended to the case in which ${\cal N}$ is not flat.
We thus conclude that
the correct interpretation is given by the Cauchy-Fueter equations, since
they can be fully generalized to the triholomorphicity condition that we
defined in the introduction.

There are three equivalent ways of defining holomorphic functions. One
requirement is differentiability
in the complex sense, another is analyticity in the complex sense
and the third definition is provided by the Cauchy-Riemann equations.
The corresponding definitions of triholomorphic functions are not
equivalent. As a matter of fact,
we know, from the mathematical literature on the subject \cite{sudbery},
that the Cauchy-Fueter equations are
the best definition of triholomorphic maps.
The other
possible ways of generalizing the definition of holomorphic maps,
are either too restrictive or too general.
The requirement of differentiability in the quaternionic sense
is too restrictive, because only the linear functions satisfy this condition
\cite{sudbery}.
The other possibility, that is analyticity in the quaternionic sense,
is too general, because it turns out that all real analytic functions in
four variables are analytic in the quaternionic sense.
The simple argument that proves it is
that, if $x=Ix_1+Jx_2+Kx_3+x_4$, then one has
\begin{equation}
\matrix{
x_1={1\over 4}(-Ix-xI-KxJ+JxK),&
x_2={1\over 4}(-Jx+KxI-xJ-IxK),\cr
x_3={1\over 4}(-Kx-JxI+IxJ-xK),&
x_4={1\over 4}(x-IxI-JxJ-KxK).}
\end{equation}
It is clear that any power series in
${x_1,x_2,x_3,x_4}$, is a power series in $x\in {\bf H}$.
We thus accept the Cauchy-Fueter equations as the best definition of
triholomorphic
functions. Triholomorphic functions will be, in general, solutions to
the Cauchy-Fueter equations, while hyperinstantons will be those solutions
that are defined globally.

Some properties of holomorphic functions are extended to triholomorphic
ones, some others are not. We shall briefly recall the main
properties to give the
reader an idea
of what triholomorphic functions are.

The complex conjugate $\bar x$ of $x$ is defined to be
$\bar x=-Ix_1-Jx_2-Kx_3+x_4$.
The norm of $x$ is $|x|=\sqrt{x \bar x}=\sqrt{x_1^2+x_2^2+x_3^2+x_4^2}$.

Notice that $q(x)=x$ is not a solution to (\ref{cauchyfueter}), while
$q(x)=Ix$, $q(x)=Jx$ and $q(x)=Kx$ are solutions.
As a matter of fact, one can easily show that all the linear solutions
are of the form
\begin{equation}
q(x)=Ix\alpha_1+Jx \alpha_2+Kx \alpha_3+\beta,
\label{sol}
\end{equation}
where $\alpha_1,\alpha_2,\alpha_3,\beta\in {\bf H}$.

The analogue
of the Cauchy formula is the Cauchy-Fueter formula \cite{sudbery}
\begin{equation}
q(x)={1\over 2\pi^2}\int_{\partial \Omega}
{x^\prime-x\over |x^\prime-x|^4}Dx^\prime q(x^\prime),
\end{equation}
where $\Omega$ is an open set containing $x$ and such that $q$ is
triholomorphic in an open set containing $\Omega$.
The three-form $Dx$ is defined as \cite{sudbery}
$Dx=dx_1\wedge dx_2\wedge dx_3+i dx_4\wedge dx_2 \wedge dx_3+
j dx_4\wedge dx_3\wedge dx_1+k dx_4\wedge dx_1\wedge dx_2$.

Any triholomorphic function is harmonic \cite{sudbery,gentili}, because
\begin{equation}
\partial\bar \partial=\bar \partial\partial=\Delta,
\end{equation}
$\Delta$ being the Laplacian.
Hence the maximum modulus principle holds. The harmonic
character of triholomorphic functions
is a consequence of the general fact that
any solution to the hyperinstanton equations
is also a solution to the equations of motion.

A theorem about analyticity holds \cite{sudbery}.
It states that any triholomorphic function can be expanded in power series
in the quaternionic numbers
(but not the most general power series) with a well-defined convergence
radius.
For the details and
other properties of the Cauchy-Fueter equations, the reader should
look at the mathematical literature.

The above definition and the above properties of triholomorphic functions
hold for the flat case ${\bf H}$.
An interesting problem would be to generalize the Cauchy-Fueter formula
and the analyticity theorem to the triholomorphic maps that we defined
with equation (\ref{afeq}) in the introduction.

We conclude this section by mentioning a
couple of straightforward properties of the general
equations (\ref{afeq}) or (\ref{inst1})
that are simple generalizations of the corresponding properties of
holomorphic functions. First of all, any constant
function is trivially triholomorphic.
{}From the point of view of topological $\sigma$-models,
the constant map
corresponds to mapping the world-manifold into a single
point of the target manifold. The moduli space of these hyperinstantons
is clearly the target manifold itself and, if the
formal dimension is positive (in which case we expect it to be
equal to the real dimension), then the physical amplitudes
are the intersection forms of the target manifold. The only nonvanishing
observables are the local ones. These features are common to
topological $\sigma$-models in two dimensions.

Secondly, there exists no triholomorphic function
whose image is one-dimensional, i.e.\ a curve in the target manifold.
This corresponds to the property
that any holomorphic function is an open mapping
(and so its image cannot be a curve in ${\bf C}$). Indeed,
let $q=q(t)$ be a curve in the target manifold, such that there exists
a map $t(x)$ from an open subset of the world-manifold
into it. Then we write $\partial_\mu q^i=t_\mu \dot q^i$, where the dot
denotes the derivative with respect to $t$ and $t_\mu=\partial_\mu t$.
Equations (\ref{afeq2}) give
\begin{equation}
t_\mu\dot q^i-{(j_u)_\mu}^\nu t_\nu \dot q^j {(J_u)_j}^i=0.
\label{urk}
\end{equation}
These equations and the hermiticity of the ${\cal M}$- and ${\cal N}$-metrics
$g_{\mu\nu}$ and $h_{ij}$, respectively, imply
\begin{equation}
0=g^{\mu\nu}h_{ij}[t_\mu\dot q^i-{(j_u)_\mu}^\rho t_\rho \dot q^k
{(J_u)_k}^i]\,
[t_\nu\dot q^j-{(j_v)_\nu}^\sigma t_\sigma \dot q^l {(J_v)_l}^j]=4t^2 \dot q^2,
\end{equation}
where $t^2=g^{\mu\nu}t_\mu t_\nu$ and $\dot q^2=h_{ij}\dot q^i\dot q^j$.
Consequently, either $t_\mu=0$ or $\dot q^i=0$. In both cases $q$
is the constant map.
The property that we have now proven
will be confirmed explicitly in the case of the torus, where it will
be also shown that solutions mapping the world-manifold into
two-, three- or four-dimensional
submanifolds of the target manifold can exist.

\section{Hyperinstantons on the torus}
\label{torus}

In this section we consider the case in which the world-manifold
and the target manifold are four-tori. We find all the solutions to the
hyperinstanton equations. Moreover, at the end of the section we
exhibit some examples of hyperinstantons of $S^4$ and ${\bf CP}^2$.

The torus $T_4$ is described by four quaternionic numbers $a_i$,
$T_4={\bf H}/\sim$, where, if $x,x^\prime$ denote points of
${\bf H}$, the equivalence relation $\sim$
is defined according to
\begin{equation}
x\sim x^\prime \,\,\, {\rm if}\,\, \, \exists\,n_1,n_2,n_3,n_4\in {\bf Z}
\,\,\, {\rm such \,\,\, that}\,\,\, x-x^\prime=n_ia_i.
\end{equation}
We adopt the following convention. The index of $a_i$ enumerates the four
quaternionic numbers that describe the torus. When we want to specify the
components
of $a_i$ as a vector of ${\bf R}^4$, we introduce a second index
according to $a_i=a_{ij}e_j$, where $(e_1,e_2,e_3,e_4)=(I,J,K,1)$.
Thus $a_{ij}$ denotes the $j$-th component of the $i$-th vector. The
torus $T_4$ is described by the matrix $a\equiv (a_{ij})$.
In fact $a$ is the period matrix of the torus.
We shall not
distinguish the indices $\mu$ of the world-manifold from the indices $i$
of the target. Moreover, all indices will be lower indices.

We look for the solutions $q:T_4\rightarrow T^\prime_4$ to the equation
$\bar\partial q=0$. $T_4^\prime$ will be described by quaternionic numbers
$b_i$ instead of $a_i$. We shall write $T_4^\prime={\bf H}/\approx$.
For $T_4^\prime$ the analogous of the matrix $a=(a_{ij})$ will be
denoted by $b=(b_{ij})$.

We must impose $q(x)\approx q(x^\prime)$ if $x\sim x^\prime$.
This requirement implies that $q$ is linear in $x$, as we now prove.
Consider the partial
derivatives $\partial_i q_j$ of $q$. We have $\partial_i q_j
(x+n_ka_k)=\partial_i q_j(x)$. The partial derivatives, being periodic,
can be expanded in
Fourier series, and consequently the functions $q_i(x)$ are the sums of
linear functions plus Fourier series
\begin{equation}
q_i=\beta_i+x_j\alpha_{ji}+\sum_{n=(n_j)\in{\bf Z}^4-\{0\}}
c_i(n)\,{\rm e}^{2\pi ix^ta^{-1}n}.
\label{piri}
\end{equation}
The reality condition gives $c^*_i(n)=c_i(-n)$, as usual.
The partial derivatives
\begin{equation}
\partial_i q_j=\alpha_{ij}+2\pi i\sum_{n=(n_j)\in{\bf Z}^4}
(a^{-1})_{il}n_l c_j(n)\,{\rm e}^{2\pi ix^ta^{-1}n}
\end{equation}
must satisfy equation (\ref{flat}). For the coefficients $c_i(n)$ this means,
if we put $f_i^{(n)}=(a^{-1})_{ik}n_k$,
\begin{equation}
\matrix{f_i^{(n)} c_i=0,&
f_{[i}^{(n)}c_{j]^+}=0.}
\end{equation}
These are four equations in four unknowns. The solution is trivial since
the determinant
of the matrix of coefficients is $f_i^{(n)}f_i^{(n)}=
n^t(a^{-1})^ta^{-1}n>0$ for $n\neq 0$.
We conclude that $q$ is linear in $x$.
The theorem that we have just proved is the four dimensional
counterpart of an analogous theorem holding for holomorphic maps between
two-tori: the derivative of the map is not only holomorphic, but also periodic
and this forces it to be constant, because there does not exist a nontrivial
bounded holomorphic function (in our case the derivative
of the map) defined all over ${\bf C}$. Thus the map is forced to be linear.

The topological field theory under consideration is a free theory, namely the
lagrangian
has the form
\begin{equation}
{\cal L}=\partial_\mu q^i\partial_\mu q^i+\bar \zeta_I\partial\!\!\!\slash
\zeta^I.
\end{equation}
Consequently, the formal dimension of the moduli space is zero.
In the partition function the bosonic and fermionic contributions factorize.
Due to the zero modes of the fermions (that are independent of the particular
hyperinstanton) the fermion factor integrates to zero. We shall only
be interested in the bosonic factor of the partition function (which
we shall call ``partition function", for simplicity). This function
collects in a simple expression a lot of information about hyperinstantons.

Among all the possible maps $q:T_4\rightarrow T_4^\prime$,
that can be grouped according to homotopy classes,
the topological gauge-fixing (i.e.\ the condition of triholomorphicity)
picks up special representatives (hyperinstantons) in the homotopy classes.
The preferred representatives are linear in our theory.
We recall from the previous section that the only linear
triholomorphic maps are
of the form (\ref{sol}). It remains to study what restrictions are
to be imposed on
the tori in order to have nontrivial solutions. The condition is
$\exists \alpha_1,\alpha_2,\alpha_3\in {\bf H}$ such that
$\forall n=(n_i)\in{\bf Z}^4$ $\exists m(n)=(m_i(n))\in{\bf Z}^4$ such that
$I n_i a_i \alpha_1+Jn_i a_i \alpha_2+K n_i a_i \alpha_3=m_i(n)b_i$.
A necessary and sufficient condition can be obtained by choosing $n_i=v_i$,
where $v_i$ are the unit vectors of ${\bf R}^4$ (and so also of ${\bf Z}^4$),
namely there must exist a matrix of integers $m_{ij}$ and three quaternionic
numbers $\alpha_1,\alpha_2,\alpha_3$, such that
\begin{equation}
Ia_i\alpha_1+Ja_i\alpha_2+Ka_i\alpha_3=m_{ij}b_j.
\label{solfa}
\end{equation}
There are twelve unknowns
(the components
of $\alpha_1,\alpha_2,\alpha_3\in {\bf H}$) and sixteen equations. It follows
that condition (\ref{solfa}) is nontrivial. Two tori $T_4$ and $T_4^\prime$
that satisfy a condition like (\ref{solfa})
for nontrivial $\alpha$'s and $m$ will be called {\sl commensurable
tori} and condition (\ref{solfa}) will be called {\sl
commensurability condition}.

In the notation $q_i=\beta_i+x_j\alpha_{ji}$ the commensurability condition
reads
\begin{equation}
a\alpha=mb.
\label{reci}
\end{equation}
Solving $\alpha=a^{-1}mb$, the hyperinstanton condition (\ref{flat}) reads
\begin{equation}
\matrix{
{\rm tr} \, a^{-1}mb=0,&
(a^{-1}mb)_{[ij]^+}=0.}
\label{hypercond}
\end{equation}

Given $a_i,b_i\in{\bf H}$,
the solutions (if any) to the commensurability condition are discrete,
due to the presence of the matrix
of integers $m$. We conclude that the only moduli of the
hyperinstantons of the torus are the translation modes $\beta$ of equation
(\ref{sol}), i.e.\ the moduli space is $T_4^\prime$.
It is clear that the space of solutions $\alpha$ to the condition
$a\alpha b^{-1}\in {\bf Z}^{4\times 4}$ is a lattice, so that any linear
integer
combination of solutions is a solution. Let $g$ be its dimension ($g\leq 12$)
and $\alpha_n$, $n=1,\ldots g$ be a set of generators. $g$ will be called
{\sl hyperinstanton dimension}. We can write $\alpha=m_n\alpha_n$,
$m_n\in {\bf Z}$.

The bosonic action, suitably normalized, is equal to
\begin{equation}
{\cal S}={1\over 2}{\rm tr}\, [\alpha\alpha^t]=
{1\over 2}{\rm tr}[a^{-1}mbb^tm^t(a^{-1})^t]=
{1\over 2}m_nm_k{\rm tr}\,[\alpha_n\alpha^t_k]=
M^tGM,
\end{equation}
where $M$ is the vector $(m_n)$ and $G$ is the matrix $(G_{nk})$ with
\begin{equation}
G_{nk}={1\over 2}{\rm tr}\, [\alpha_n\alpha^t_k].
\end{equation}
The partition function (bosonic factor) is then a theta function,
\begin{equation}
Z=\Theta(G)=\sum_{M\in{\bf Z}^g}{\rm e}^{-M^tGM}.
\label{piola}
\end{equation}
Concluding,
the hyperinstanton problem associates to a couple of tori, $T_4$
and $T_4^\prime$,
a third torus, $T_4^{\prime\prime}$,
the dimension of which (i.e.\
the hyperinstanton dimension) takes values between
zero and twelve. Given a torus, say $T_4$, described by the matrix $a$,
one can associate to it the theta function $\Theta(aa^t/2)$, since
the matrix $aa^t$ is positive definite (the factor $1/2$ is a convention).
Correspondingly, the theta function $\Theta(bb^t/2)$ is
associated to $T_4^\prime$. One easily verifies that the theta function
associated to $T_4^{\prime\prime}$ is the partition function $Z$.
Indeed, let us expand $\alpha_n$ in a basis of $g$ matrices $E_k$
which are orthonormal with respect to the canonical scalar product
$<E_kE_l>={\rm tr}\, [E_kE_l^t]$. Then, if
$\alpha_n=\alpha_{nk}E_k$, the torus $T_4^{\prime\prime}$ is
described by the period matrix $\alpha_{nk}$ and the associated theta
function is $\Theta(\alpha\alpha^t/2)=Z$.

Let us consider the simple case
$a_i=a e_i$ and $b_i=b e_i$ for certain
$a,b$ real and positive.
The hyperinstantons are then
$q_i=\beta_i+x_j\alpha_{ij},$
where
\begin{equation}
\alpha={b\over a}\left(\matrix{
n_1&m_1+p_1&m_2+p_2&m_3+p_3\cr
-m_1+p_1&n_2&-m_3+p_4&m_2+p_5\cr
-m_2+p_2&m_3+p_4&n_3&-m_1+p_6\cr
-m_3+p_3&-m_2+p_5&m_1+p_6&-n_1-n_2-n_3}\right),
\label{matrix}
\end{equation}
with $n_j,m_j,p_j\in {\bf Z}$.
As one can see, there is no solution that maps the world-torus
into a one dimensional submanifold of the target torus. On the contrary there
are solutions whose image are two-, three- or four-dimensional.
For example, in the first case, take $n_1\neq 0$ and all the rest zero;
in the second case, take $n_1,n_2\neq 0$
and all the rest zero;
in the third case take $n_1,n_2,n_3\neq 0$
and all the rest zero.

The partition function (\ref{piola}) is in this case
\begin{equation}
Z(t)=[\Theta(t)]^9\,
\Theta\left({t\over 2}A\right),
\label{ju}
\end{equation}
where $t=b^2/a^2$ and $A$ denotes the $3\times 3$ matrix
\begin{equation}
A=\left(\matrix{2&1&1\cr 1&2&1\cr 1&1&2}\right).
\end{equation}
As mentioned in the introduction, an interesting problem (the generalization
of the problem of counting the number of rational curves of a K\"ahler
manifold) is to count the number of triholomorphic embeddings of
$S^4,{\bf CP}^2,T_4$ and K3 into hyperK\"ahler or quaternionic manifolds.
The partition function that we have just computed gives the answer in the
case of triholomorphic embeddings of a torus $T_4$ into a
second torus $T_4^\prime$. If we write
\begin{equation}
Z(t)=\sum_{n\in {\bf N}}m(n)\,{\rm e}^{-nt},
\end{equation}
then $m(n)$ represents the number of triholomorphic embeddings such that
the topological invariant (\ref{stac}) or (\ref{st0})
(which is {\sl not} the winding number) takes the value $n$.

Let us exhibit another example.
It is a case in which the hyperinstanton dimension in neither
twelve nor zero. Let $a={\rm diag}\, (1,1,v,z)$ with $v$ and $z$ irrational
and such that there exists no vanishing integer linear combination
of $v,z,vz$. Let $b=\zeta \cdot 1$, with $\zeta\in {\bf R}$. Then one can
show that the most general matrix $m$ is
\begin{equation}
m=\zeta^{-1}\alpha=\left(\matrix{n_1&n_2&n_3&n_4\cr n_2&-n_1&n_4&-n_3\cr
0&0&0&0\cr 0&0&0&0}\right),
\label{hy}
\end{equation}
where $n_1,n_2,n_3,n_4\in{\bf Z}$,
so that the hyperinstanton dimension is $g=4$ and the partition function is
\begin{equation}
Z(\zeta^2)=[\Theta(\zeta^2)]^4.
\end{equation}
Choosing $a={\rm diag}\, (1,u,v,z)$
with $u$, $v$ and $z$ irrational
and such that there exists no vanishing integer
linear combination of $uvz,uv,uz,vz$,
we get a case in which
the hyperinstanton dimension is zero. The above examples were chosen so as to
justify the name commensurability condition that we gave to
condition (\ref{solfa}).

We now study the deformations ${\cal D}^{\alpha_p,\beta_k}_{p,k}$
(\ref{mostgen}) of the topological theory.
We shall restrict to the case ${\cal D}^{\alpha_{4-k},\beta_k}_{4-k,k}$,
that is the case when the deformations
contain no ghosts, to preserve the factorization between bosonic and fermionic
integrations.
Then the deformed (bosonic) partition function is of the general form
\begin{eqnarray}
\phantom{.}&&
Z(Q^i_1,Q^{jk}_2,Q^{lmn}_3,Q_4^{pqrs})\nonumber\\
\phantom{.}&&
\equiv\sum_{N=(N_i)\in {\bf Z}^g}{\rm exp}\,
\{Q_1^iN_i+Q_2^{jk}N_jN_k+Q_3^{lmn}N_lN_mN_n+
Q_4^{pqrs}N_pN_qN_rN_s\}.
\end{eqnarray}
Functions of this form
will be called {\sl hyper-theta-functions}. They have the property
\begin{eqnarray}
\phantom{>}&&Z(Q^i_1,Q^{jk}_2,Q^{lmn}_3,Q_4^{pqrs})=\nonumber\\
\phantom{.}&&{\rm exp}\,
\{Q_1^iM_i+Q_2^{jk}M_jM_k+Q_3^{lmn}M_lM_mM_n+
Q_4^{pqrs}M_pM_qM_rM_s\}\nonumber\\
\phantom{.}&&\cdot Z(Q^i_1+2Q^{ij}_2M_j+3
Q^{ijk}_3M_jM_k+4Q_4^{ijkl}M_jM_kM_l,\nonumber\\
\phantom{.}&&
Q^{jk}_2+3Q^{jkl}_3M_l+6Q_4^{jklm}M_lM_m,
Q^{lmn}_3+4Q_4^{lmnp}M_p,Q_4^{pqrs}),
\end{eqnarray}
$\forall M\in {\bf Z}^g$. This property
expresses the fact that under suitable deformations
the partition function changes in a very simple way, namely it is
multiplied by a factor. Hyper-theta-functions with $Q_3=Q_4=0$
are the usual theta-functions. Since the winding number is quartic
in the integers $N_i$, we see that the number of triholomorphic embeddings
of $T_4$ into $T_4^\prime$ with a given winding number is encoded
into a hyper-theta-function, precisely as the number of triholomorphic
embeddings with a given value of the topological invariant
(\ref{stac}) or (\ref{st0}) is encoded into a theta-function
like (\ref{piola}).

We conclude this section with some remarks about the case ${\cal M}=
{\cal N}=S^4$ and  the case ${\cal M}={\cal N}={\bf CP}^2$.  Let us begin
with ${\cal M}={\cal N}=S^4$.
Using stereographic coordinates, the hyperinstanton equations have
the form (\ref{flat}) is both the northern and the southern chart. Let
$x$ and $q$ be quaternionic numbers referring
to the northern charts of ${\cal M}$ and
${\cal N}$, respectively, and $x^\prime=x/|x|^2$,
$q^\prime=q/|q|^2$ be the southern coordinates. Differently from
the case of the torus, the triholomorphic functions are not subject
to periodicity conditions, rather to the (nontrivial) condition that
they should satisfy the hyperinstanton equations in both charts.
For instance, the
map $q(x)=(r_1 I+r_2 J +r_3 K)x\alpha$, where $r_1,r_2,r_3\in {\bf R}$
and $\alpha\in {\bf H}$, is a good solution.
$r_1,r_2,r_3$ and $\alpha$ are moduli. Indeed, when $x$ approaches the
southern pole of ${\cal M}$, then $q$ also approaches the southern pole
of ${\cal N}$. Changing charts in both cases, we find
$q^\prime(x^\prime)
=(r^\prime_1 I+r^\prime_2 J +r^\prime_3 K)x^\prime \alpha^\prime$,
where $r_j^\prime=r_j/(r_1^2+r_2^2+r_3^2)$ and $\alpha^\prime=\alpha/|\alpha|^2
$,
which is clearly a solution to the southern equations.
These solutions have winding number one.
We do not know
what is the complete set of solutions $q:S^4\rightarrow S^4$.
It would be certainly interesting, to be compared with the
two-dimensional analogue of the holomorphic functions $f:S^2\rightarrow S^2$.
As a matter of fact, $S^2$ is also ${\bf CP}^1$, so that the best
four dimensional counterpart could well be the set of
triholomorphic maps
$q:{\bf CP}^2\rightarrow {\bf CP}^2$. As an example, writing the equations
in a suitable relative Lorentz gauge, the identity map
${\bf CP}^2\rightarrow {\bf CP}^2$ is easily shown to be
a solution.
Other interesting possibilities are the embeddings of $S^4$ or
${\bf CP}^2$ into the Grassmannian $SO(m+4)/(SO(4)\otimes SO(m))$.

\section{Hyperinstantons on the K3 surface}
\label{k3}

Now we analyze the case in which both the world-manifold and the target
manifold are
the K3 surface and search for hyperinstantons. We first discuss the
hyperinstanton equations
in the form (\ref{inst1}).
We suppose that
the world-manifold satisfies $\omega^{-ab}=0$, i.e.\ we choose the
suitable Lorentz reference frame in which this happens (this is required
when writing the hyperinstanton equations in the form (\ref{inst1}),
that comes from the topological twist).

The first attempt that one tries is to see if the identity map is a
hyperinstanton. In order for this to be true, we have to use a trick, because
there is an obstruction. Looking back to the first of equations
(\ref{flat}), one is lead to infer that the identity map is not a solution,
since the equation $V_a^\mu E_i^a\partial_\mu q^i=0$ of (\ref{inst1}),
that generalizes the flat equation
$\partial_\mu q_\mu=0$ of (\ref{flat}), would not be satisfied.
However, a suitable trick can make it a solution.

We choose the target vierbein $E_i^a(q)$ equal, in form, to the
world-manifold vierbein $V_\mu^a(x)$, apart from a global Lorentz rotation
$\Lambda^{ab}$
(the relative Lorentz gauge of section \ref{physics}), namely we
set
\begin{equation}
E_i^a(q)={\Lambda^a}_b V^b_\mu(q)\delta^\mu_i.
\label{vierbein}
\end{equation}
Equations (\ref{inst1}) then become
\begin{equation}
\matrix{V^\mu_a(x){\Lambda^a}_b V^b_\rho(q(x))\delta^\rho_i
\partial_\mu q^i(x)=0,\cr
V^{\mu [a}(x){\Lambda^{b]^+}}_c V^c_\rho(q(x))
\delta^\rho_i\partial_\mu q^i(x)=0.}
\end{equation}
It is then clear that the identity map
$q:K3\rightarrow K3$, $q=x$ is a solution,
provided the orthogonal matrix $\Lambda^{ab}$ is traceless and its
antisymmetric part is antiselfdual
\begin{equation}
\matrix{{\Lambda^a}_a=0,&\Lambda^{[ab]^+}=0.}
\label{puro}
\end{equation}
For example, we can choose $\Lambda^{ab}=I_1^{ab}$,
with $I_1$ given in equation (\ref{opuwe}) (in fact any
matrix $I_u$ of (\ref{opuwe}) is good for this purpose).

Notice that the identity map
is always a solution of the equations of motion when the four dimensional
target manifold
is the same as the world-manifold.
The equation of motion
\begin{equation}
\partial_\mu(\sqrt{g(x)}g^{\mu\nu}(x)\partial_\nu q^i(x))
+\sqrt{g(x)}\Gamma^i_{jk}(q(x))\partial_\mu q^j(x)
\partial_\nu q^k(x) g^{\mu\nu}(x)=0
\end{equation}
is surely satisfied by $q^i(x)=x^\mu\delta_\mu^i$,
if $h_{ij}(q)=g_{\mu\nu}(q)\delta^\mu_i\delta^\nu_j$, as follows from
(\ref{vierbein}),
because it reduces to the identity
\begin{equation}
\partial_\mu(\sqrt{g}g^{\mu\nu})
+\sqrt{g}\Gamma^\nu_{\rho\sigma}g^{\rho\sigma}=0.
\label{identity}
\end{equation}
With some more computational effort, one can check that
any isometry is a solution to the field equations. The natural question
is whether any isometry is also solution to the hyperinstanton equations and
what classification between isometries is induced by the hyperinstanton
equations that is not induced by the field equations.

An isometry is
a map $q:K3\rightarrow K3$ such that
\begin{equation}
g_{\rho\sigma}(q(x))\delta^\rho_i\delta^\sigma_j
\partial_\mu q^i(x)\partial_\nu q^j(x)=g_{\mu\nu}(x).
\end{equation}
This means that there must exist an orthogonal matrix ${A^a}_b(x)$
such that
\begin{equation}
V_\rho^a(q(x))\delta^\rho_i\partial_\mu q^i={A^a}_b(x)V^b_\mu(x).
\end{equation}
In this case, the hyperinstanton equations reduce to
\begin{equation}
\matrix{{\Lambda^a}_b{A^b}_a=0,&
(\Lambda A)^{[ab]^+}=0.}
\end{equation}
It is clear that it is sufficient to choose ${\Lambda^a}_b={(I_1)^a}_c{(A^{-1})
^c}_b$. This shows that for any given isometry
there exists a relative Lorentz gauge such that
the hyperinstanton equations are solved.
However, we must notice that not all these solutions contribute to the same
topological field theory. As we noticed at the end of section
\ref{physics}, the relative Lorentz gauge must be chosen
once for all, because ${\Lambda^a}_b$ enters in the expression (\ref{vierbein})
of the vierbein of the target manifold. This is the classification between
isometries induced by the hyperinstanton equations.

It is simple to see that, if the target K3 manifold is rescaled with respect
to the world-manifold, then the above isometries can be turned into rescalings,
thus giving a set of solutions that is in one-to-one correspondence
with the above set of isometries.

Before going on with the argument, let us reexamine the above
reasoning on equations (\ref{afeq}). To make the identity
map a solution, we use the ambiguity in the formulation of the equations
(related to the concept of relative Lorentz gauge, as already pointed out),
to put them into the form (\ref{afeq4}), namely
\begin{equation}
q_*-\Lambda^{uv} \, J_u\circ q_* \circ j_v=0,
\label{afeq3}
\end{equation}
where $\Lambda^{uv}$ is an $SO(3)$ matrix that parametrizes the ambiguity.
We choose $\Lambda^{uv}\equiv\Lambda_0^{uv}=
{\rm diag}\, (1,-1,-1)$ and, since
${\cal M}={\cal N}$, we can set $J_u=j_u$, so that the identity
map satisfies (\ref{afeq3}), since
\begin{equation}
1-j_1^2+j_2^2+j_3^2=0.
\end{equation}
It is clear that in general not all isometries satisfy (\ref{afeq3})
with the above $\Lambda_0^{uv}$. We can derive how many of them do
satisfy (\ref{afeq3}) with the chosen $\Lambda_0^{uv}$, by
making use of the results by Alekseevsky \cite{alekseevsky}.

It is well known that K3 possesses no conformal Killing vector, so the
set of isometries of  K3 is a discrete one.

In fact, K3 possesses lots of isometries\footnotemark\footnotetext{We are
grateful to
M.\ Pontecorvo and D.V.\ Alekseevsky and to S.\ Cecotti about this point.}
and there is a simple argument for finding them \cite{alekseevsky}.
For simplicity, we shall limit ourselves to consider K3
as the Fermat surface $F$ \cite{taormina} in ${\bf CP}^3$
\begin{equation}
X_1^4+X_2^4+X_3^4+X_4^4=0.
\label{fermat}
\end{equation}
The embedding of K3 in ${\bf CP}^3$ induces a K\"ahler metric $\kappa$.
$\kappa$ in not
the Calabi-Yau metric, of course, however the Calabi-Yau metric $g$ is defined
as the unique
Ricci-flat K\"ahler metric $g$ whose K\"ahler form is
cohomologus to the K\"ahler form of the metric
$\kappa$. Let $G$ be the set of
holomorphic transformations of ${\bf CP}^3$ that preserve the surface
(\ref{fermat}).
These transformations leave $\kappa$ invariant. Since $g$ is
uniquely determined
by $\kappa$, $g$ is invariant under $G$. ${G}$ is thus a set of
isometries of the
Calabi-Yau metric $g$.
One finds \cite{alekseevsky} $G=S_4\cdot ({\bf Z}_4)^3$, where
$S_4$ represents the permutation group of the four homogeneous
${\bf CP}^3$-coordinates $X_1$, $X_2$, $X_3$ and $X_4$, while ${\bf Z}_4$
represents the possibility of multiplying them by the fourth roots
of the identity $1,i,-1,-i$. There are only three ${\bf Z}_4$'s and not
four, because the overall one is immaterial.
One can then consider the set $G^\prime=\tau G$ of antiholomorphic
transformations
of ${\bf CP}^3$ that preserve $F$, where $\tau$ denotes the
antiholomorphic
involution of
$F\subset {\bf CP}^3$ defined by complex conjugation of the homogeneous
coordinates
of ${\bf CP}^3$. These transformations are also isometries.
In ref.\ \cite{alekseevsky} it is also
proven that
any isometry is of one of the two types $G,\tau G$ that we have described.
So, the full group of isometries of $g$ turns out to
be $G\cup\tau G$, which contains $2^{10}\cdot 3=3072$ elements.

Let $E$ be the space of parallel two-forms on the Fermat surface,
$E\approx {\bf R}^3$.
Define the following scalar product on $E$. Given $\alpha,\beta\in E$,
let $\alpha\cdot \beta$ be equal to
\begin{equation}
\alpha\cdot\beta=\int_{K3}\alpha\wedge \beta.
\end{equation}
Alekseevsky shows that an isometry $q$ induces a rotation on $E$.
The set of isometries is a discrete group, so it has a natural representation
as a finite
subgroup of $SO(3)$ acting on $E$. Since the complex structures
contracted with the metric span $E$, an isometry has
got the same effect on the set of complex structures as on the space $E$.
Let $R_q^{uv}$ be the rotation induced by $q$ on $E$.
We have
\begin{equation}
q_*\circ j_u=R_q^{uv}\, j_v \circ q_*.
\label{utility}
\end{equation}
Then
$q$ satisfies (\ref{afeq3}), if we choose, for example,
$\Lambda= \Lambda_0\cdot
(R_q)^{-1}$.

Let $G_0$ be the set of isometries $q$ that belong to the kernel
of the representation on $E$ (i.e.\ such that $R_q=1$).
$G_0$ is determined by Alekseevsky \cite{alekseevsky}
and can be characterized
in the following way. Represent an isometry
of $G\cup \tau G$ by a $4\times 4$ complex matrix acting on
the vector $(X_1,X_2,X_3,X_4)$ in ${\bf C}^4$.
The determinant of this matrix can only take
the values $1,i,-1,-i$. $G_0$ is the set of isometries such that
this determinant is one. It is a normal subgroup of $G\cup \tau G$ and
contains $2^7\cdot 3=384$ elements. Moreover,
$G\cup\tau G$ acts on $E\approx {\bf R}^3$ as the dyhedral group $D_4$.
Precisely,
\begin{equation}
{G\cup\tau G\over G_0}\approx D_4.
\label{6.14}
\end{equation}
Thus, $R_q\in D_4\subset SO(3)$. We can
restrict the matrices $\Lambda$ of
eq.\ (\ref{afeq3}) to be also in $D_4\subset SO(3)$,
since, for any given isometry $q$
there exists a $\Lambda\in D_4$ such that $q$ solves (\ref{afeq3}).
In some sense,
the group $D_4$ ``measures" the ambiguity
in the condition of triholomorphicity.

Using (\ref{utility}), it is immediate to see
that an isometry $q$ solves eq.\ (\ref{afeq3}) if and only if
${\rm tr}\, \Lambda R_q=-1$ and $\Lambda R_q$ is
symmetric ($\Lambda_0$ is an example of such
symmetric $D_4$-matrices with trace $-1$).
By inspection in the eight matrices of $D_4$, one checks that there
are five such $R_q$'s. This number is independent of $\Lambda\in D_4$, of
course, but the set of ``good'' $R_q$'s does depend on $\Lambda$.
In conclusion, due to (\ref{6.14}),
the total number of hyperinstantonic isometries $q: K3\rightarrow K3$ is
$384\cdot 5=1920$, whatever $\Lambda\in D_4$ we choose, but the set
of hyperinstantonic isometries depends on the chosen $\Lambda$.

We have thus shown that the hyperinstanton equations induce an
interesting structure in the group of isometries, that does not follow
from the field equations. Moreover, we have characterized the solutions by
simple properties of the Fermat surface. We thus address the possibility
that the condition of triholomorphicity has a purely algebraic formulation
that applies to the cases of algebraic varieties.

To conclude this section, we prove that isometries are isolated
hyperinstantons in
the moduli space, namely that equation (\ref{zeromodes})
admits no nontrivial solution. Equation (\ref{zeromodes})
is simplified by the fact that ${\cal D}_k{(J_u)_j}^i=0$
for K3. Moreover, it
must be adapted
to our choice of the relative Lorentz gauge, so we substitute it with
\begin{equation}
{\cal D}_\mu\xi^i-\Lambda^{uv}{(j_u)_\mu}^\nu
{\cal D}_\nu\xi^j {(J_v)_j}^i=0.
\label{mode}
\end{equation}
Let $\xi^\mu$ be defined by $\xi^i=\xi^\mu\partial_\mu q^i$. Using the identity
${\cal D}_\mu \xi^i={\cal D}_\mu\xi^\nu\,
\partial_\nu q^i$ and eq.\ (\ref{utility}) with $\Lambda=
\tilde\Lambda (R_q)^{-1}$
where $\tilde\Lambda$ is any symmetric $D_4$-matrix with trace $-1$,
any explicit $q$-dependence disappears from (\ref{mode})
and we get
\begin{equation}
{\cal D}_\mu\xi^\nu-\tilde\Lambda^{uv}{(j_u)_\mu}^\rho
{\cal D}_\rho\xi^\sigma {(j_v)_\sigma}^\nu=0.
\end{equation}
Using a standard argument, we show that the solutions $\xi^\mu$
of the above equation are Killing vectors. Since K3 admits no
Killing vector, there are no solutions. In fact,
with some standard manipulations such as integration by parts
and the use of the self-duality of the Riemann tensor (we are thinking
of K3 with the Calabi-Yau metric) and the covariant
constancy of the three complex structures ${(j_u)_\mu}^\nu$, one proves
the following identity
\begin{eqnarray}
0&=&\int_{K3}d^4 x \, \sqrt{g} g^{\mu \alpha}g_{\nu\beta}
({\cal D}_\mu\xi^\nu-\tilde\Lambda^{uv}{(j_u)_\mu}^\rho
{\cal D}_\rho\xi^\sigma {(j_v)_\sigma}^\nu)
({\cal D}_\alpha\xi^\beta-\tilde\Lambda^{st}{(j_s)_\alpha}^\gamma
{\cal D}_\gamma\xi^\zeta{(j_t)_\zeta}^\beta)\nonumber\\
&=&4\int_{K3}d^4 x \, \sqrt{g}g^{\mu \alpha}g_{\nu\beta}
{\cal D}_\mu\xi^\nu{\cal D}_\alpha\xi^\beta.
\end{eqnarray}
This shows that $\xi^\nu$ must necessarily satisfy ${\cal D}_\mu\xi^\nu=0$,
so it vanishes. The formal dimension of the moduli-space turns out to be
negative, since one easily checks that the topological antighosts do
possess zero-modes
(the constants).

We notice that the set of hyperinstantons of the K3 surface that we have
exhibited may be not the whole set of hyperinstantons. The complete
identification of K3 hyperinstantons remains an open problem. A convenient
way towards this classification is possibly offered by the analysis
of the algebraic counterpart of the hyperinstanton equations.

All the above arguments can be extended, with the obvious modifications, to the
cases when K3 is described by other algebraic surfaces than the Fermat surface.

\section{Hyperinstantons with dynamical gravity}
\label{dynamical}

In this section we discuss the case in which gravity is dynamical
and we illustrate the difficulty inherent to the problem of solving
the coupled hyperinstanton equations (\ref{inst1}) and (\ref{inst2}),
of which no nontrivial solution has so far been found.
In particular, we show that the simplest  possibilities,
namely the identity map
between identical four dimensional quaternionic manifolds,
cannot be turned into solutions to the hyperinstanton equations.
The same maps are good solutions for the topological $\sigma$-model
formulated in section \ref{sigmamodel}.

Let the target manifold be any quaternionic ${\cal N}$.
The trivial solution is when
$q({\cal M})$ is a point in $\cal N$. This means $q=$constant, so that
the hyperinstanton equations (\ref{inst1}) are satisfied provided
the world-manifold ${\cal M}$ is hyperK\"ahler (${\cal M}=T_4$ or
${\cal M}=K3$ if it is compact).
Clearly, the moduli space of these solutions is the entire target
manifold ${\cal N}$ times the moduli space of the world-manifold
${\cal M}$.

{}From now on, we concentrate on four dimensional ${\cal N}$
(in four dimensions, the only quaternionic
manifolds are $S^4$ and
${\bf CP}^2$, by definition).
The first reasonable guess would be to conjecture
that the solutions that we have
found in the case of K3 have a counterpart in the present case. In particular
one might think that the identity map\footnotemark
\footnotetext{It is sufficient to study the identity map.
Now diffeomorphisms are
a gauge symmetry and they should be gauge-fixed. Thus, under
suitable gauge-fixing, only one
diffeomorphism matters. We choose it to be the identity.}
and the world-manifold
equal to the target manifold (or equal to it up
to a rescaling)
is a solution to the hyperinstantonic equations.
This is not the case. The trick that has been
successful in solving the last two equations of (\ref{inst1})
is ruled out by the first one.

Before showing this,
let us consider a different problem, namely a generic
$\sigma$-model coupled to dynamical gravity such that the target
manifold is four dimensional.
We can write ${\cal M}={\cal M}({\cal N})$
and $q=q({\cal N})$, to mean that ${\cal N}$ is given, while
$\cal M$ and $q$ must be determined
(to be precise, ${\cal M}$ is given as a topological space and its
metric is the unknown). We do not restrict
$\cal N$ to be a quaternionic manifold, for now. We wonder what are
the {\sl eigenmanifolds} ${\cal M}({\cal N})={\cal N}$, so that
the diffeomorphisms $q:{\cal N}\rightarrow {\cal N}$ are solutions.
The Euclidean lagrangian is
\begin{equation}
{1\over 2}
\sqrt{g} (-R+\lambda g^{\mu\nu}h_{ij}(q)\partial_\mu q^i\partial_\nu q^j).
\label{pirite}
\end{equation}
The equations of motion are
\begin{eqnarray}
0&=&R_{\mu\nu}-{1\over 2}g_{\mu\nu}R-T_{\mu\nu},\nonumber\\
0&=&\partial_\mu(\sqrt{g(x)}g^{\mu\nu}(x)\partial_\nu q^i(x))
+\sqrt{g(x)}\Gamma^i_{jk}(q(x))\partial_\mu q^j(x)
\partial_\nu q^k(x) g^{\mu\nu}(x),
\label{equations}
\end{eqnarray}
where the energy-momentum tensor $T_{\mu\nu}$ turns out to be
\begin{equation}
T_{\mu\nu}=\lambda h_{ij}(q)\partial_\mu q^i\partial_\nu q^j
-{1\over 2} \lambda g_{\mu\nu}g^{\rho\sigma}h_{ij}(q)\partial_\rho q^i
\partial_\sigma q^j.
\end{equation}

Let us consider the identity
map $q:{\cal N}\rightarrow {\cal N}$. Then
the first equation of (\ref{equations}) gives
\begin{equation}
R_{\mu\nu}=\lambda g_{\mu\nu},
\label{poi}
\end{equation}
so that the target manifold is forced to be Einstein, with
cosmological constant equal to $\lambda$. On the other hand,
the second equation of (\ref{equations}) is surely satisfied,
in force of the identity (\ref{identity})
that we have already used in the previous section. Thus we have proved that
any Einstein manifold is an eigenmanifold
(if the cosmological constant is equal to $\lambda$).
In view of the invariance
under diffeomorphisms, we can extend the conclusion to
any diffeomorphism $q:{\cal N}\rightarrow {\cal N}$.

Notice that the cosmological constant of the target manifold is forced
to be equal to the parameter $\lambda$ that appears in (\ref{pirite}).
One can find more general solutions if a cosmological term $\lambda_0\sqrt{g}$
is added to (\ref{pirite}) \cite{iengo}, but we are not interested
to this case, since it cannot come from the topological
twist of N=2 supergravity coupled to matter and
a twist-independent formulation of the
topological $\sigma$-model coupled to topological gravity is still missing.

If ${\cal M}$ is chosen to be equal to a rescaling of ${\cal N}$, then
the identity map $q=x$ can be replaced by $q=\xi x$, $\xi$ being the rescaling
factor. As before, one easily checks that $q=\xi x$ is a solution
only if ${\cal N}$ is Einstein and its
cosmological constant of ${\cal N}$ is equal to $\lambda$.

Now, the theory of  N=2 supergravity coupled to hypermultiplets
\cite{dauriaferrarafre,baggerwitten} gives another relation between the
cosmological constant $\Lambda$ of ${\cal N}$ and the factor $\lambda$
appearing in the lagrangian (\ref{pirite}), namely
$\Lambda=3\lambda$ (in general,
$\Lambda=\lambda (n^2+2n)$, if ${\rm dim}\, {\cal N}=4n$),
so that the above simple ansatz are not even solutions to
the field equations of the theory that comes from the twist,
and {\sl a fortiori} they are not hyperinstantons.
We now prove this fact on the hyperinstanton equations themselves,
to illustrate where the trick that was successful for K3 fails.

Let us consider the identity map $q:{\cal N}\rightarrow {\cal N}$.
We already know how to satisfy equations
(\ref{inst1}). It is sufficient to impose (\ref{vierbein})
(with a $\Lambda^{ab}$ that can now depend on the point)
and to choose a relative Lorentz gauge
$\Lambda^{ab}$ such that ${\Lambda^a}_a=0$ and
$\Lambda^{[ab]^+}=0$. Then, the problem is to satisfy
(\ref{inst2}). With $q^i=\delta^i_\mu x^\mu$ and
$E_i^a(q)={\Lambda^a}_b V^b_\mu (q) \delta^\mu_i$, this
equation becomes
\begin{equation}
\omega^{-ab}={\cal L}_\Lambda \omega^{-ab},
\label{rota}
\end{equation}
where ${\cal L}_\Lambda$ denotes the Lorentz transformation
performed by the orthogonal matrix $\Lambda^{ab}$.
In other words, there must exist a traceless Lorentz transformation
such that its antisymmetric part is antiselfdual and to which the
antiselfdual part of the spin connection is insensitive. We can easily prove
that this cannot be. Let us suppose that eq.\ (\ref{rota})
is satisfied. Then we would have
\begin{equation}
R^{-ab}={\cal L}_\Lambda R^{-ab}.
\end{equation}
Expanding $R^{-ab}$ in a basis of anti-self-dual matrices $I_u^{ab}$
like the ones of (\ref{opuwe}),
$R^{-ab}=I_u^{ab}R^u$, we conclude $[\Lambda,I_u]=0$ for any
$u$ such that $R_u\neq 0$. Since all $R_u$ are different from zero,
because the target manifold has $SU(2)$ right-holonomy, $\Lambda$
must be proportional to the identity. This is absurd, since the trace of
$\Lambda$ should be zero.

In conclusion, a nontrivial example of hyperinstanton with dynamical gravity
is so far still missing, due to the difficulty of the coupled equations.
So, the eventual solutions are surely very peculiar ones and could
exhibit very interesting properties. A much simpler problem
is given by the case of quaternionic world and target manifolds in the
external gravity regime, as shown in section \ref{sigmamodel}.
The maps that we have considered in this section are indeed hyperinstantons
of the  topological $\sigma$-model of section \ref{sigmamodel}
(see the end of section \ref{torus}).

\section{Conclusions}

There are two ways of formulating a topological field theory:
the first is by topologically twisting an N=2 supersymmetric theory and
the second is by BRST-quantizing
a suitable continuous deformation as a gauge-symmetry and
imposing  ``by hand'' a suitable instantonic condition as a gauge-fixing.
The second method is more general, since not all topological field theories
can be obtained by the twist procedure.
On the other hand, the topological twist has the advantage
that it automatically yelds a gauge-fixed topological field theory.
In this paper, we combined the properties
of the two methods to get the most general formulation of  topological
$\sigma$-models in four dimensions. We first took advantage of the
topological twist in order to get a hint of the so far unknown
instantonic equations. Secondly, we searched for their most general
mathematical
interpretation and we found that they are a condition
of triholomorphicity on the map. Finally, we went back and formulated
the most general topological $\sigma$-model
by BRST-quantizing the continuous deformations of the map and
imposing the  triholomorphicity condition
as a gauge-fixing.
So, the spirit of our investigation was not
the search for a reformulation of
a mathematical problem and of mathematical results
in the language of physics.
Rather, we formulated a new
mathematical problem inspired by a physical theory.

Indeed, the identification of topological $\sigma$-models in four dimensions
with the concept of triholomorphic maps proposes the study of a quite
interesting class of mappings on which very few results are so far
known in the mathematical literature, i.e.\ the triholomorphic
embeddings of four dimensional Riemannian manifolds into almost
quaternionic manifolds.
{}From the physical
point of view, some subtleties that are not obvious in the
two-dimensional case
are also revealed by our analysis of topological field theories in four
dimensions. More generally, the problem of
coupling topological $\sigma$-models to topological gravity (a problem
whose two-dimensional analogue was solved in terms of classical integrable
hierarchies) is shown by our work to be related to an even less
studied mathematical problem, namely
that of hyperinstantons consistently coupled to gravitational instantons
\cite{anselmifre}.
The difficulties inherent to the solutions of the coupled differential
equations have been illustrated in our work.

\vspace{24pt}
\begin{center}
{\bf Acknowledgements}
\end{center}
We would like to thank G.\ Gentili, S.\ Vidussi, M.\ Pontecorvo,
D.V.\ Alekseevsky, S.\ Salamon,
S.\ Cecotti and C.N.\ Pope for helpful discussions.
\vspace{24pt}

\appendix

\section{HyperK\"ahler and quaternionic manifolds}
\label{hyperappendix}

In this appendix we give definitions and properties of a quaternionic manifold
${\cal Q}(m)$. We shall mention the changes that occur for
${\cal Q}(m)$ hyperK\"ahler. In any case,
the formul\ae\ for a hyperK\"ahlek manifold can be retrieved
by formally substituting the symbol $\Omega_u$ with $\lambda \Omega_u$
everywhere, simplifying the $\lambda$'s
whenever possible and then
letting $\lambda$ go to zero. Moreover, $\omega^u$ are set to zero.

${\cal Q}(m)$ is first of all
a $4m$-dimensional Riemannian manifold. We denote its metric by
\begin{equation}
ds^2=h_{ij}(q) dq^i\otimes dq^j.
\end{equation}
Moreover, ${\cal Q}(m)$ possesses
an qlmost quaternionic structure, namely three locally defined \cite{galicki}
$(1,1)$-tensors
$J^u$, $u=1,2,3$, fulfilling the quaternionic algebra
\begin{equation}
J^u J^v= - \delta^{uv}+\varepsilon^{uvz}J^z.
\label{complex}
\end{equation}
The metric $h_{ij}$ is Hermitian with respect to all the
almost quaternionic $(1,1)$-tensors.
$J_u$ are indeed globally defined and covariantly constant
complex structures
if ${\cal Q}(m)$ is hyperK\"ahler.

We introduce the three forms
\begin{equation}
\Omega^u=\lambda h_{ik}{(J^u)_j}^kdq^i\wedge dq^j.
\label{2.56}
\end{equation}
$\lambda$ is a real constant that is related to the cosmological constant of
${\cal Q}(m)$. Indeed,  any quaternionic manifold is an Einstein manifold.

In the hyperK\"ahler case, the forms $\Omega^u$ are K\"ahler forms,
namely
\begin{equation}
d\Omega^u=0.
\end{equation}
If ${\cal Q}(m)$ is quaternionic, there exist three one-forms $\omega^u$ that
make an $SU(2)$ connection, with respect to which the forms $\Omega^u$
are covariantly closed and such that $\Omega^u$ is the field strength
of this connection, namely
\begin{eqnarray}
d\Omega_u+\varepsilon_{uvz}\omega^v\wedge\Omega^z&=&0,\nonumber\\
d\omega_u+{1\over 2}\varepsilon_{uvz}\omega^v\wedge\omega^z&=&\Omega_u.
\end{eqnarray}

The general feature of ${\cal Q}(m)$ is that
its holonomy group $Hol({\cal Q}(m))$ is contained in $SU(2)\otimes
Sp(m)$\footnotemark\footnotetext{In our notation, $Sp(m)$ denotes the
symplectic group in $2m$ dimensions.}.
In the hyperK\"ahler case, the $SU(2)$ part of the spin connection
of ${\cal Q}(m)$ is flat, while in the quaternionic case
its curvature is proportional to $\Omega^u$,
where $h_{ij}$ is the metric of ${\cal Q}(m)$.

We can introduce a quaternionic vielbein
${\cal U}^{A I}_i$ where $A=1,2$ is an index of $SU(2)$
and $I=1,\ldots 2m$ is an index of $Sp(m)$.
Let us introduce the vielbein one form
\begin{equation}
{\cal U}^{A I}={\cal U}^{A I}_i dq^i.
\label{viel}
\end{equation}
We have
\begin{equation}
h_{ij}={\cal U}^{A I}_i {\cal U}^{B J}_j {\cal C}_{IJ}\varepsilon_{AB},
\label{2.60}
\end{equation}
where ${\cal C}_{IJ}$
is the flat $Sp(m)$ invariant metric, while $\varepsilon_{AB}$
is, of course, the flat $Sp(1)\equiv SU(2)$ flat invariant metric.

The vielbein ${\cal U}^{A I}$ is covariantly closed with respect
to the $SU(2)$-connection $\omega^u$ and to some $sp(m)$-valued
connection $\Delta^{IJ} = \Delta^{JI}$, namely
\begin{eqnarray}
\nabla {\cal U}^{AI}&\equiv& d{\cal U}^{AI} -{i\over 2} \omega^u
(\varepsilon \sigma_u\varepsilon^{-1})^A_{\phantom{A}B}\wedge{\cal U}^{BI}
\nonumber\\
&&+ \Delta^{IJ} \wedge {\cal U}^{AK} {\cal C}_{JK}=0,
\label{de}
\end{eqnarray}
where $(\sigma^x)_A^{\phantom{A}B}$ are the standard Pauli matrices.
Furthermore ${
\cal U}^{AI}$ satisfies  the reality condition
\begin{equation}
{\cal U}_{AI} \equiv ({\cal U}^{AI})^* = \varepsilon_{AB}
{\cal C}_{IJ} {\cal U}^{BJ}.
\label{2.62}
\end{equation}
Eq.(\ref{2.62})  defines  the  rule to lower the symplectic indices by
means   of  the  flat  symplectic   metrics   $\varepsilon_{AB}$   and
${\cal C}_{IJ}$.

We also introduce the inverse vielbein ${\cal U}^i_{AI}$ defined by the
equation
\begin{equation}
{\cal U}^i_{AI} {\cal U}^{AI}_j = \delta^i_j.
\end{equation}

Flattening a pair of indices of the Riemann  tensor
${\cal R}^{ij}_{\phantom{uv}{kl}}$
we obtain
\begin{equation}
R^{ij}_{\phantom{uv}{st}} {\cal U}^{AI}_i {\cal U}^{BJ}_j = \Omega^u_{st}
 {i\over 2} (\varepsilon^{-1} \sigma_u)^{AB} {\cal C}^{IJ}
+ {{\cal R}^{IJ}}_{st}
\varepsilon^{AB},
\label{2.65}
\end{equation}
where ${{\cal R}^{IJ}}_{st}$ is the field strength of the $Sp(m)$
connection
\begin{equation}
d \Delta^{IJ} + \Delta^{IK} \wedge \Delta^{LJ}
{\cal C}_{KL} \equiv {\cal R}^{IJ} = {{\cal R}^{IJ}}_{st}
dq^s \wedge dq^t.
\end{equation}
Eq. (\ref{2.65}) is the explicit statement that the Levi Civita connection
associated with the metric $h$ has a holonomy group contained in
$SU(2) \otimes Sp(m)$. Consider now eq.s (\ref{complex}) and (\ref{2.56}).
We easily derive the following relation
\begin{equation}
h^{st} \Omega^u_{is} \Omega^v_{tj} = - \lambda^2\delta^{uv} h_{ij} +
\lambda\varepsilon^{uvz} \Omega^z_{ij}.
\label{2.67}
\end{equation}
Eq. (\ref{2.67}) implies that the intrinsic components of the 2-form
$\Omega^u$ yield a representation of the quaternionic algebra. Hence we
can set
\begin{equation}
\Omega^u_{AI,BJ} \equiv \Omega^u_{ij} {\cal U}^i_{AI}
{\cal U}^j_{BJ} = - i \lambda {\cal C}_{IJ} (\sigma^u
\varepsilon)_{AB}.
\label{2.68}
\end{equation}
Alternatively eq. (\ref{2.68}) can be rewritten in an intrinsic form as
\begin{equation}
\Omega^u = i \lambda {\cal C}_{IJ} (\sigma^u \varepsilon^{-1})_{AB}
{\cal U}^{AI} \wedge {\cal U}^{BJ},
\end{equation}
wherefrom we also get
\begin{equation}
{i\over 2} \Omega^u (\sigma_u)_A^{\phantom{A}B} =
\lambda{\cal U}_{AI} \wedge {\cal U}^{BI}.
\label{kalviel}
\end{equation}

\section{The topological twist}
\label{twist}

In this appendix we briefly recall the most general description of
hypermultiplets \cite{baggerwitten,dauriaferrarafre},
also coupled to N=2 supergravity and we perform the BRST
quantization. Then we proceed to define the topological twist and to show
that the gauge-fixing equations are those that we expect. The main
steps, without details, were given in ref.\ \cite{anselmifre2}.

To fix the notation, we denote the hypermultiplets by
$(q^i,\zeta_I,\zeta^I)$.
$q^i$ are the
coordinates of the 4$m$-dimensional target manifold ${\cal Q}(m)$
($i=1,\ldots 4m$). $\zeta_I$ and
$\zeta^I$ are the left
handed and right handed components of the fermionic superpartners
($I=1,\ldots 2m$),

Specifically, ${\cal Q}(m)$ is a hyperK\"ahler manifold when gravity
is external, while it is a quaternionic manifold
when gravity is dynamical (i.e.\ the hypermultiplets are coupled to
supergravity). In the first case one should put
restrictions on the gravitational
background in order to have global supersymmmetry. However, we
know that the topological theory is meaningful
in a more general background. We focus on hypermultiplets coupled
to supergravity, since the case of N=2 global supersymmetry can be retrieved
as a suitable limit of the N=2 locally supersymmetric theory.

The parameter
$\lambda$ of (\ref{2.56}) appears in front of the kinetic term
in the action of the hypermultiplets
\cite{dauriaferrarafre} [see also formula (\ref{lagra})].
In order to get a physical kinetic term in the case of dynamical
gravity, the sign of $\lambda$ should be fixed and
the target manifold should be
noncompact (when the signature is Minkowskian)
\cite{baggerwitten}. However, we shall not put this restriction,
since the complete action is anyway nonpositive definite (due to
the Einstein term). Moreover, the topological version of
the theory is well defined in itself and the action
is zero on any solution to the field
equations and {\sl a fortiori} on any instantonic solution.
As we know, any quaternionic manifold is Einstein.
Precisely, in the Minkowskian signature,
we have ${\cal R}_{ij}=-(m^2+2m)\lambda h_{ij}$.

The ``generalized curvatures" are the one forms (\ref{viel}) and the following
covariant derivatives of the fermions $\zeta_I$ and $\zeta^I$
\begin{eqnarray}
\nabla \zeta_I=d\zeta_I-{1\over 4}\gamma_{ab}\omega^{ab}\wedge
\zeta_I+{\Delta_I}^J\zeta_J={\cal D}\zeta_I+{\Delta_I}^J\zeta_J,\nonumber\\
\nabla \zeta^I=d\zeta^I-{1\over 4}\gamma_{ab}\omega^{ab}\wedge
\zeta^I-{\Delta^I}_J\zeta^J={\cal D}\zeta^I-{\Delta^I}_J\zeta^J,
\label{curv}
\end{eqnarray}
where $\omega^{ab}$ is the world-manifold Lorentz spin connection while
${\Delta_I}^J$ is the $Sp(m)$ connection
${\Delta_I}^J={\cal C}_{IK}\Delta^{KJ}$, ${\Delta^I}_J={\cal C}_{jK}
\Delta^{IK}$.

The superspace rheonomic parametrizations of the generalized curvatures are
easily found \cite{dauriaferrarafre}:
\begin{eqnarray}
{\cal U}^{AI}&=&{\cal U}_a^{A I}V^a+
\varepsilon^{AB}{\cal C}^{IJ}
\bar\psi_B\zeta_J+\bar \psi^A\zeta^I,\nonumber\\
\nabla \zeta_I&=&\nabla_a\zeta_IV^a+i{\cal U}^{BJ}_a\gamma^a\psi^A
\varepsilon_{AB}{\cal C}_{IJ},\nonumber\\
\nabla \zeta^I&=&\nabla_a\zeta^IV^a+i{\cal U}^{AI}_a\gamma^a\psi_A.
\label{rheo}
\end{eqnarray}

We also report the definitions and
the rheonomic
parametrizations of the curvatures of N=2 supergravity, which are
necessary in the case ${\cal Q}(m)$ is quaternionic.
The definitions are
\begin{eqnarray}
R^a&=&dV^a-\omega^{ab}\wedge V_b-i\bar\psi_A\wedge \gamma^a
\psi^A={\cal D}V^a-i\bar\psi_A\wedge \gamma^a
\psi^A,\nonumber\\
R^{ab}&=&d\omega^{ab}-{\omega^a}_c\wedge \omega^{cb},\nonumber\\
\rho_A&=&d\psi_A-{1\over 4}\gamma_{ab}\omega^{ab}\wedge\psi_A+
{\omega_A}^B\wedge \psi_B={\cal D}\psi_A+
{\omega_A}^B\wedge \psi_B,\nonumber\\
\rho^A&=&d\psi^A-{1\over 4}\gamma_{ab}\omega^{ab}\wedge\psi^A+
{\omega^A}_B\wedge \psi^B={\cal D}\psi^A+
{\omega^A}_B\wedge \psi^B,\nonumber\\
F&=&dA+\bar\psi_A\wedge \psi_B\varepsilon^{AB}+
\bar\psi^A\wedge \psi^B\varepsilon_{AB},
\label{curv2}
\end{eqnarray}
where $\psi_A$ and $\psi^A$ are the left handed and right handed
components of the gravitinos, respectively, while
${\omega_A}^B=i{1\over 2}{(\sigma_u)_A}^B\omega^u$
and ${\omega^A}_B=\varepsilon^{AL}{\omega_L}^M\varepsilon_{MB}$.

The rheonomic parametrizations are
\begin{eqnarray}
R^a&=&0,\nonumber\\
R^{ab}&=&{R^{ab}}_{cd}V^c\wedge V^d+
\varepsilon^{abcd}\bar\psi^A\wedge\gamma_d\psi_B({A^B}_{A|c}-
{{\bar A}^B}_{\phantom{.}A|c})
\nonumber\\&&
-i\bar\psi_A (2\gamma^{[a}
\rho^{A|b]c}-\gamma^c\rho^{A|ab})\wedge V_c-i
\bar\psi^A(2\gamma^{[a}
\rho_A^{b]c}-\gamma^c\rho_A^{ab})\wedge V_c
\nonumber\\&&
-\varepsilon^{AB}\bar\psi_A\wedge \psi_B F^{-ab}
-\varepsilon_{AB}\bar\psi^A\wedge\psi^B F^{+ab},\nonumber\\
F&=&F_{ab}V^a\wedge V^b,\nonumber\\
\rho_A&=&\rho_{A|ab}V^a\wedge V^b+{A^B}_{A|a}\gamma^{ab}\psi_B
\wedge V_b+i\varepsilon_{AB}F^{+ab}
\gamma_a\psi^B\wedge V_b,\nonumber\\
\rho^A&=&\rho^A_{|ab}V^a\wedge V^b+{{\bar A}^A}_{\phantom{.}B|a}\gamma^{ab}
\psi^B
\wedge V_b+i\varepsilon^{AB}F^{-ab}
\gamma^a\psi_B\wedge V^b,\nonumber\\
\label{rheo2}
\end{eqnarray}
where
\begin{equation}
{A^B}_{A|a}={1\over 4}\lambda\delta_A^B\bar\zeta^I\gamma_a\zeta_I.
\end{equation}

Now we give the kinetic lagrangian of N=2 supergravity coupled to
hypermultiplets \cite{dauriaferrarafre}.
In the case ${\cal Q}(m)$ is hyperK\"ahler, the lagrangian
is formally the same, upon suppression of the terms involving gravitinos
and the Einstein kinetic term and upon setting $\omega^u$ equal to zero.
\begin{eqnarray}
{\cal L}_{kin}&=&\varepsilon_{abcd} R^{ab} \wedge V^c\wedge V^d
-4(\bar\psi^A\wedge \gamma_a\rho_A-\bar\psi_A\wedge\gamma_a
\rho^A)\wedge V^a\nonumber\\
&&-{4\over 3}\lambda\varepsilon_{AB}{\cal C}_{IJ}{\cal U}^{AI}_a({\cal U}^{BJ}-
\bar \psi^B\zeta^J-\varepsilon^{BC}{\cal C}^{JK}\bar\psi_C\zeta_K)
\wedge V_b\wedge V_c\wedge V_d \varepsilon^{abcd}\nonumber\\
&&+i{2\over 3}\lambda(\bar\zeta^I\gamma_a\nabla\zeta_I+
\bar\zeta_I\gamma_a\nabla\zeta^I)
\wedge V_b\wedge V_c\wedge V_d \varepsilon^{abcd}\nonumber\\
&&+{1\over 6}\lambda{\cal U}^{AI}_e{\cal U}^{e \ BJ}
\varepsilon_{AB}{\cal C}_{IJ}
V_a\wedge V_b\wedge V_c\wedge V_d \varepsilon^{abcd}\nonumber\\
&&+{1\over 12}F^{ab}F_{ab}\,\varepsilon_{cdef}V^c\wedge V^d\wedge
V^e\wedge V^f-\varepsilon_{abcd}F^{ab}V^c\wedge V^d\wedge F.
\label{lagra}
\end{eqnarray}
We now perform the BRST quantization of the theory. We follow
the general procedure described in refs.\
\cite{baulieubellon,anselmifre,anselmifre2}.
The local symmetries
involved are diffeomorphisms, Lorentz rotations, the gauge symmetry
related to the graviphoton $A$ and supersymmetries. The ghosts will be
denoted by $\epsilon^a$, $\epsilon^{ab}$, $c$ and $c_A$, $c^A$,
respectively.
One finds
\begin{eqnarray}
sq^i&=&{\cal U}_{AI}^i({\cal U}_a^{A I}\epsilon^a+
\varepsilon^{AB}{\cal C}^{IJ}
\bar c_B\zeta_J+\bar c^A\zeta^I),\nonumber\\
s\zeta_I&=&{1\over 4}\gamma_{ab}\epsilon^{ab}\wedge
\zeta_I-{\Delta_I}^J_{(0,1)}\zeta_J+
\nabla_a\zeta_I\,\epsilon^a+i{\cal U}^{BJ}_a\gamma^a c^A
\varepsilon_{AB}{\cal C}_{IJ},\nonumber\\
s\zeta^I&=&{1\over 4}\gamma_{ab}\epsilon^{ab}\wedge
\zeta^I+{{\Delta^I}_J}_{(0,1)}\zeta^J+
\nabla_a\zeta^I\epsilon^a+i{\cal U}^{AI}_a\gamma^a c_A,
\nonumber\\
sV^a&=&-{\cal D}\epsilon^a+
\epsilon^{ab}\wedge V_b
+i\bar c_A\wedge \gamma^a\psi^A+i\bar\psi_A\wedge \gamma^a
c^A,\nonumber\\
s\epsilon^a&=&\epsilon^{ab}\wedge \epsilon_b+i\bar c_A\wedge \gamma^a
c^A,\nonumber\\
s\omega^{ab}&=&-{\cal D}\epsilon^{ab}
+2{R^{ab}}_{cd}\epsilon^c\wedge V^d
-i(\bar c_A \wedge V_c+\bar\psi_A\epsilon_c)(2\gamma^{[a}
\rho^{A|b]c}-\gamma^c\rho^{A|ab})\nonumber\\&&-i(\bar c^A\wedge V_c+
\bar\psi^A\epsilon_c)(2\gamma^{[a}
\rho_A^{b]c}-\gamma^c\rho_A^{ab})
\nonumber\\&&+
\varepsilon^{abcd}(\bar c^A\wedge\gamma_d\psi_B+
\bar\psi^A\wedge\gamma_d c_B)
({A^B}_{A|c}-{{\bar A}^B}_{\phantom{.}A|c})
\nonumber\\&&
-\varepsilon^{AB}(\bar c_A\wedge \psi_B
+\bar\psi_A\wedge c_B)F^{-ab}
-\varepsilon_{AB}(\bar c^A\wedge\psi^B+\bar\psi^A\wedge c^B)
F^{+ab},\nonumber\\
s\epsilon^{ab}&=&{\epsilon^a}_c\wedge \epsilon^{cb}+
{R^{ab}}_{cd}\epsilon^c\wedge \epsilon^d-i\bar c_A (2\gamma^{[a}
\rho^{A|b]c}-\gamma^c\rho^{A|ab})\wedge\epsilon_c\nonumber\\&&
-i\bar c^A(2\gamma^{[a}
\rho_A^{b]c}-\gamma^c\rho_A^{ab})\wedge \epsilon^c+
\varepsilon^{abcd}\bar c^A\wedge\gamma_d c_B({A^B}_{A|c}-
{{\bar A}^B}_{\phantom{.}A|c})
\nonumber\\&&-\varepsilon^{AB}\bar c_A\wedge c_B F^{-ab}
-\varepsilon_{AB}\bar c^A\wedge c^BF^{+ab},\nonumber\\
s\psi_A&=&-{\cal D}c_A
+{1\over 4}\gamma_{ab}\epsilon^{ab}\wedge\psi_A-
{\omega_A}^B\wedge c_B-
{\omega_A}^B_{(0,1)}\wedge \psi_B\nonumber\\&&+
2\rho_{A|ab}\epsilon^a\wedge V^b+{A^B}_{A|a}\gamma^{ab}
(c_B\wedge V_b+\psi_B\wedge \epsilon_b)\nonumber\\&&
+i\varepsilon_{AB}F^{+ab}
\gamma_a(c^B\wedge V_b+\psi^B\wedge \epsilon_b),\nonumber\\
sc_A&=&{1\over 4}\gamma_{ab}\epsilon^{ab}\wedge c_A-
{\omega_A}^B_{(0,1)}\wedge c_B+
\rho_{A|ab}\epsilon^a\wedge \epsilon^b+{A^B}_{A|a}\gamma^{ab}c_B
\wedge \epsilon_b\nonumber\\&&
+i\varepsilon_{AB}F^{+ab}
\gamma_a c^B\wedge \epsilon_b,\nonumber\\
s\psi^A&=&-{\cal D}\psi^A
+{1\over 4}\gamma_{ab}\epsilon^{ab}\wedge\psi^A-{\omega^A}_B\wedge c^B-
{{\omega^A}_B}_{(0,1)}\wedge \psi^B
\nonumber\\&&+2\rho^{A|ab}\epsilon_a\wedge V_b+
{{\bar A}^A}_{\phantom{.}B|a}\gamma^{ab}
(c^B\wedge V_b+\psi^B\wedge \epsilon_b)\nonumber\\&&
+i\varepsilon^{AB}F^{-ab}
\gamma^a(c_B\wedge V^b+\psi_B\wedge \epsilon^b),\nonumber\\
sc^A&=&{1\over 4}\gamma_{ab}\epsilon^{ab}\wedge c^A-
{{\omega^A}_B}_{(0,1)}\wedge c^B+\rho^{A|ab}\epsilon_a\wedge \epsilon_b
+{{\bar A}^A}_{\phantom{.}B|a}\gamma^{ab}c^B
\wedge \epsilon_b\nonumber\\&&
+i\varepsilon^{AB}F^{-ab}
\gamma^a c_B\wedge \epsilon^b,\nonumber\\
sA&=&-dc-2\bar c_A\wedge \psi_B\varepsilon^{AB}-2
\bar c^A\wedge \psi^B\varepsilon_{AB}+2F_{ab}\epsilon^a\wedge V^b,
\nonumber\\
sc&=&-\bar c_A\wedge c_B\varepsilon^{AB}-
\bar c^A\wedge c^B\varepsilon_{AB}+F_{ab}\epsilon^a\wedge \epsilon^b,
\label{brst}
\end{eqnarray}
where $\Delta^{IJ}_{(0,1)}$ and ${{\omega^A}_B}_{(0,1)}$ are obtained from
the one forms $\Delta^{IJ}$ and ${\omega^A}_B$ upon substitution of the
differential $dq^i$ with the BRST variation $sq^i$, which appears in the first
equation of (\ref{brst}).

Now we perform the topological twist of N=2 supergravity
coupled to hypermultiplets.
As mentioned in appendix \ref{hyperappendix}, $Hol({\cal Q}(m))\subset
SU(2)\otimes Sp(m)$.
This $SU(2)$ is the internal
supersymmetry automorphism $SU(2)_I$ \cite{dauriaferrarafre,anselmifre2}.
In the
twisted version of the theory, $SU(2)_I$ is identified
with the right handed part $SU(2)_R$ of the Lorentz group, to give
the new $SU(2)_R^\prime$.
We recall \cite{anselmifre2} that
one also has to identify an $SU(2)_Q$
for the redefinition
of $SU(2)_L$: it is the $SU(2)$ factor in
the $SU(2)\otimes SO(m)$ maximal subgroup of $Sp(m)$.
Moreover, there must exist a suitable internal $U(1)$,
in order to perform the redefinition of the ghost number.
This $U(1)_I$ is R-duality \cite{anselmifre2}.

The complete twisting procedure can be divided in
three steps. Step A corresponds to the redefinitions of $SU(2)_L$,
$SU(2)_R$
and ghost number $U(1)_g$ according to the following scheme
\begin{eqnarray}
SU(2)_L&\longrightarrow & SU(2)_L^\prime={\rm diag}
[SU(2)_L\otimes SU(2)_Q],\nonumber\\
SU(2)_R&\longrightarrow & SU(2)_R^\prime={\rm diag}
[SU(2)_R\otimes SU(2)_I],\nonumber\\
U(1)_g&\longrightarrow & U(1)_g^\prime={\rm diag}
[U(1)_g\otimes U(1)_I],\nonumber\\
^c(L,R,I,Q)^g_f &\longrightarrow & (L\otimes Q,R\otimes I)^{g+c}_f,
\end{eqnarray}
where $L$ denotes the representation of $SU(2)_L$,
$R$ the representation of $SU(2)_R$,
$Q$ is the representation of $SU(2)_Q$,
$c$ is the $U(1)_I$ charge, $g$ the ghost number and $f$ the form number.
$U(1)_g^\prime={\rm diag}
[U(1)_g\otimes U(1)_I]$ is intended to mean that the new ghost number
is the sum of the old ghost number plus the R-duality charge.
Step B is the correct identification
of the topological ghosts
(fields with $g+c=1$ from $g=0$, $c=1$) by contraction
with a suitable vielbein (in our case the quaternionic
vielbein ${\cal U}^{AI}_i$ for the identification
of the topological ghosts $\xi^i$ \cite{anselmifre2}).
Step C is the topological shift,
namely the shift by a constant
of the $(0,0)^0_0$-field coming by applying step A on
the right handed components
of the supersymmetry ghosts, $c^A$, namely $(c^{\dot \alpha})^A\rightarrow
-i/2 \, e \varepsilon^{\dot \alpha A}+(c^{\dot \alpha})^A$
\cite{anselmifre,anselmifre2}, where $e$ is
the broker (a zero form with fermionic statistic and ghost number one,
with the convention that $e^2$ is set equal to $1$ \cite{anselmifre2}).

The $U(1)_I$ internal symmetry (R-duality)
that adds to ghost number to give the ghost
number of the topological version of the theory is
chirality on the gravitinos, duality on the
graviphoton and the opposite of chirality on the superpartners of the
quaternionic coordinates, the hyperini $\zeta_I$ and $\zeta^I$, namely
\cite{anselmifre2}
\begin{equation}
\matrix{\hat\delta V^a=0,& \hat \delta A=0,\cr
\hat\delta \psi_A=\psi_A,&
\hat\delta \psi^A=-\psi^A,\cr
\hat\delta F^+_{ab}=2F^+_{ab},&
\hat\delta F^-_{ab}=-2F^-_{ab},\cr
\hat\delta \zeta_I=-\zeta_I,&
\hat\delta \zeta^I=\zeta^I.}
\label{rduality}
\end{equation}
R-duality extends to the BRST-quantized theory by simply stating that
any field has the same R-duality behaviour as its ghost partner.

In general, the R-duality anomaly is the formal
dimension of the moduli space, because, after the topological twist,
it represents the ghost number anomaly. The R-duality anomaly
is not only due to the axial anomaly (R-duality is proportional to chirality on
the fermions) but is also due to the dual anomaly of the graviphoton
\cite{dolgov}. This anomaly is related to the difference between the numbers
of zero modes of self-dual and anti-self-dual field strengths $F^{ab}$
\cite{reuter} and not to the zero modes of the vector $A$.
So, the problem of the ghost-number anomaly
is not so simple as in two dimensions or as in topological Yang-Mills Theory.
So far, a complete
analysis of this anomaly has not been performed.

The indices $I=1,\ldots 2m$ of $Sp(m)\supset SU(2)_Q\otimes SO(m)$
are splitted into a couple of indices, according to $I=(\alpha,k)$, where
$\alpha=1,2$ is an index of $SU(2)_Q\approx SU(2)_L$ and
$k=1,\ldots m$ is an index of $SO(m)$.
The $Sp(m)$-invariant metric ${\cal C}_{IJ}$ becomes $\varepsilon_{\alpha
\beta}\delta_{kl}$, if $I=(\alpha,k)$ and $J=(\beta,l)$.
We can introduce the vielbein $E_i^{ak}\equiv {1\over 2}
{\cal U}^{\dot A
\alpha k}_i(\sigma^a)_{\alpha \dot A}$.
Moreover,
we can define the true topological antighosts
\begin{eqnarray}
\zeta^{+ab}_k&=&-e{(\sigma^{ab})_\alpha}^\beta\varepsilon^{\alpha\gamma}
(\zeta_\beta)
_{\gamma k},\nonumber\\
\zeta_k&=&-e\varepsilon^{\alpha\beta}(\zeta_\alpha)_{\beta k},
\label{antighost}
\end{eqnarray}
which, under the new Lorentz group transform
as $(1,0)$ and $(0,0)$ respectively.

The twisted-shifted BRST algebra is, up to nonlinear terms containing ghosts,
\begin{equation}
\matrix{
sV^a = \tilde\psi^a-d\epsilon^a+\epsilon^{ab}\wedge V_b,&
s\epsilon^{a} = C^a,\cr
s\epsilon^{ab} = -{1\over 2}{F^+}^{ab},&
\cr
s\tilde\psi^a = -d C^a+{1\over 2}{F^+}^{ab}\wedge V_b,
&
s\tilde\psi^{ab} = -dC^{ab}+i{1\over 2}({\omega^-}^{ab}+{1\over 2}I_u^{ab}
q^*\omega^u),\cr
s\tilde\psi  = -dC ,&
sC^{a} = 0,\cr
sC^{ab} = {i\over 2}{\epsilon^-}^{ab}+{i\over 4}I_u^{ab}q^*\omega^u_{(0,1)},&
sC = 0,\cr
s A  =  i \tilde\psi -dc,&
sc  = -{1\over 2}+iC,\cr
sq^i = -{i\over 2}e{\cal U}^i_{\dot A I}(\zeta^{\dot A})^I\equiv \tilde\xi^i,
&
s\tilde\xi^i = 0,\cr
s\zeta^{+ab k} = -2V^{\mu[a}E_i^{b]^+k}\partial_\mu q^i,
&
s\zeta^k = V^\mu_a E^{ak}_i\partial_\mu q^i,}
\end{equation}
where the formul\ae\ relating $\tilde\psi^a,\tilde\psi^{ab},\tilde\psi$
to $\psi_A,\psi^A$ are
\begin{equation}\matrix{
\tilde\psi^a={e\over 2}(\psi_\alpha)_{\dot A} (\bar \sigma^a)^
{\dot A\alpha},&
\tilde\psi^{ab}=-e{(\bar \sigma^{ab})^{\dot A}}_{\dot \alpha}
(\psi^{\dot \alpha}) _{\dot A},&
\tilde \psi =-e(\psi_{\dot \alpha}) ^{\dot A}\delta_{\dot A}^{\dot \alpha},}
\end{equation}
and similar formul\ae\ relate $C^a,C^{ab},C$ to $c_A,c^A$. Moreover,
$[ab]^+$ means antisymmetrization and selfdualization in the indices $a,b$.
Thus we see
that {\sl both} $\zeta^{+ab}_k$ and $\zeta_k$ are topological antighosts
(otherwise we would not have enough equations to fix the gauge completely).

Let us give the gauge-free algebra
of the topological theory in full generality.
\begin{eqnarray}
sV^a&=&\psi^a-{\cal D}_0\epsilon^a+\epsilon^{ab}\wedge V_b,\nonumber\\
s\omega_0^{ab}&=&\chi^{ab}-{\cal D}_0\epsilon^{ab},\nonumber\\
s\epsilon^a&=&\phi^a+\epsilon^{ab}\wedge\epsilon_b,\nonumber\\
s\epsilon^{ab}&=&\eta^{ab}+{\epsilon^a}_c\wedge\epsilon^{cb},
\nonumber\\
s\psi^a&=&-{\cal D}_0\phi^a+\epsilon^{ab}\wedge\psi_b-\chi^{ab}\wedge
\epsilon_b-\eta^{ab}\wedge V_b,\nonumber\\
s\phi^a&=&\epsilon^{ab}\wedge\phi_b-\eta^{ab}\wedge\epsilon_b,\nonumber\\
s\chi^{ab}&=&-{\cal D}_0\eta^{ab}+\epsilon^{ac}\wedge{\chi_c}^b-\chi^{ac}
\wedge{\epsilon_c}^b,\nonumber\\
s\eta^{ab}&=&\epsilon^{ac}\wedge{\eta_c}^b-\eta^{ac}
\wedge{\epsilon_c}^b,\nonumber\\
sq^i&=&\xi^i,\nonumber\\
s\xi^i&=&0,
\label{free}
\end{eqnarray}
where the subscript $0$ means that the spin connection
$\omega^{ab}$ is the usual one (i.e.\ in the
second order formalism it does not
contain the terms quadratic in the fermions that characterize the
supergravity spin connection).
As we see, the $\sigma$-model sector of the BRST-algebra is trivial.

The complete identification of the gauge-free BRST algebra (\ref{free})
with the minimal subalgebra
of the gauge-fixed BRST algebra (\ref{brst})
is given by (the dots stand for nonlinear
terms containing ghosts)
\begin{eqnarray}
\xi^i&=&{\cal U}_{AI}^i({\cal U}_a^{A I}\epsilon^a+
\varepsilon^{AB}{\cal C}^{IJ}
\bar c_B\zeta_J+\bar c^A\zeta^I)=\tilde\xi^i+\cdots,\nonumber\\
\psi^a&=&
i\bar c_A\wedge \gamma^a\psi^A+i\bar\psi_A\wedge \gamma^a c^A
-A^{ab}\epsilon_b=
\tilde \psi^a+\cdots,\nonumber\\
\phi^a&=&i\bar c_A\wedge \gamma^ac^A=C^a+\cdots,\nonumber\\
\eta^{ab}&=&
{R^{ab}}_{cd}\epsilon^c\wedge \epsilon^d-i\bar c_A (2\gamma^{[a}
\rho^{A|b]c}-\gamma^c\rho^{A|ab})\wedge\epsilon_c-i
\bar c^A(2\gamma^{[a}
\rho_A^{b]c}-\gamma^c\rho_A^{ab})\wedge \epsilon^c\nonumber\\&&+
\varepsilon^{abcd}\bar c^A\wedge\gamma_d c_B({A^B}_{A|c}-
{{\bar A}^B}_{\phantom{.}A|c})
-\varepsilon^{AB}\bar c_A\wedge c_B F^{-ab}
-\varepsilon_{AB}\bar c^A\wedge c^BF^{+ab}\nonumber\\&&
=-{1\over 2}F^{+ab}
+\cdots.
\end{eqnarray}

The total action can be written as the sum
of the classical topological action ${\cal S}_T$ (\ref{st1})
plus the BRST variation
of the a gauge fermion $\Psi$, that, up to interaction terms containing
ghosts, turns out to be
\begin{eqnarray}
\Psi&=&-16i \left[B^{ab}-i \left(\omega^{-ab}+{1\over 2}I_u^{ab}q^*\omega^u
+2idC^{ab}\right)\right]\wedge  \psi_{ac}\wedge V_b\wedge V^c\nonumber\\
&&+8iF\wedge \psi_a \wedge V^a+{\lambda\over 3}\varepsilon_{cdef}
V^c\wedge V^d\wedge V^e\wedge V^f[2 \eta^{ab}\epsilon_{ab}
\nonumber\\&&
+(4V^{\mu[a}E_i^{b]^+k}\partial_\mu q^i+\Lambda^{abk})\zeta_{abk}
+(\Lambda^k-2 V^\mu_aE_i^{ak}\partial_\mu q^i)\zeta_k],
\end{eqnarray}
where $B^{ab}$, $\Lambda^{abk}$ and $\Lambda^k$ are
Lagrange multipliers ($s\psi^{ab}=B^{ab}$, $s\zeta^{abk}=\Lambda^{abk}$,
$s\zeta^k=\Lambda^k$, $sB^{ab}=s\Lambda^{abk}=s\Lambda^k=0$).

\end{document}